\documentclass[12pt]{article}
\usepackage[margin=1in]{geometry}
\usepackage[T1]{fontenc}
\usepackage[utf8]{inputenc}
\usepackage{graphicx}
\usepackage{amsmath,amssymb}
\usepackage{hyperref}
\usepackage[version=4]{mhchem}
\usepackage{array,tabularx,makecell} 
\usepackage{authblk}
\usepackage{orcidlink}
\usepackage[backend=biber,style=numeric,sorting=none,giveninits=true,url=true,doi=true]{biblatex}
\newcolumntype{Y}{>{\centering\arraybackslash}X} 
\newcolumntype{C}[1]{>{\centering\arraybackslash}m{#1}} 

\begin{document}

\title{A modular framework for collaborative human-AI, multi-modal and multi-beamline synchrotron experiments}

\author[1]{Adam A. Corrao\,\orcidlink{0000-0001-6111-8959}}
\author[1]{Phillip M. Maffettone\,\orcidlink{0000-0001-7173-7972}}
\author[2]{Bruce Ravel\,\orcidlink{0000-0002-4126-872X}}
\author[1]{Thomas A. Caswell\,\orcidlink{0000-0003-4692-608X}}
\author[1]{Stuart I. Campbell\,\orcidlink{0000-0001-7079-0878}}
\author[2]{Howie Joress\,\orcidlink{0000-0002-6552-2972}}
\author[1]{Stuart Wilkins\,\orcidlink{0000-0003-1191-3350}}
\author[1,*]{Daniel Olds\,\orcidlink{0000-0002-4611-4113}}

\affil[1]{National Synchrotron Light Source II, Brookhaven National Laboratory, Upton, NY, U.S.}
\affil[2]{Materials Measurement Lab, National Institute of Standards and Technology, Gaithersburg, MD, U.S.}
\affil[*]{Corresponding author: dolds@bnl.gov}

\maketitle


\begin{abstract}
    High-throughput materials discovery and studies of complex functional materials increasingly rely on multi-modal characterization performed at synchrotron light sources. However, measurements are typically done with no use of data until after an experiment, neglecting opportunities for data-driven insights to guide measurements. We developed a modular, open-source framework that incorporates artificial intelligence within the Bluesky control and data streaming infrastructure at NSLS-II, enabling real-time orchestration of multi-beamline, multi-modal experiments. AI agents perform on-the-fly reduction, clustering, Gaussian process modelling, and Bayesian optimization driven data acquisition, while users monitor agent behavior and visualize results live. Combinatorial libraries of the ternary Al-Ni-Pt system were spatially mapped by X-ray diffraction and X-ray absorption fine structure measurements at the PDF and BMM beamlines, respectively. Dynamic switching between AI-driven and conventional grid mapping strategies was achieved, demonstrating the flexible workflows possible through this framework. A digital twin constructed from a simulated Al-Li-Fe oxide dataset shows that AI-driven mapping strategies outperform conventional mapping as well as random sampling by prioritizing measurements that better resolve both phase boundaries and localized minority phases. This framework supports plug-and-play capabilities, and establishes a foundation for routine multi-modal, AI-assisted large-scale user-facility operations.
\end{abstract}

\section{Introduction}
\begin{figure}
\centering
\includegraphics[width=\textwidth]{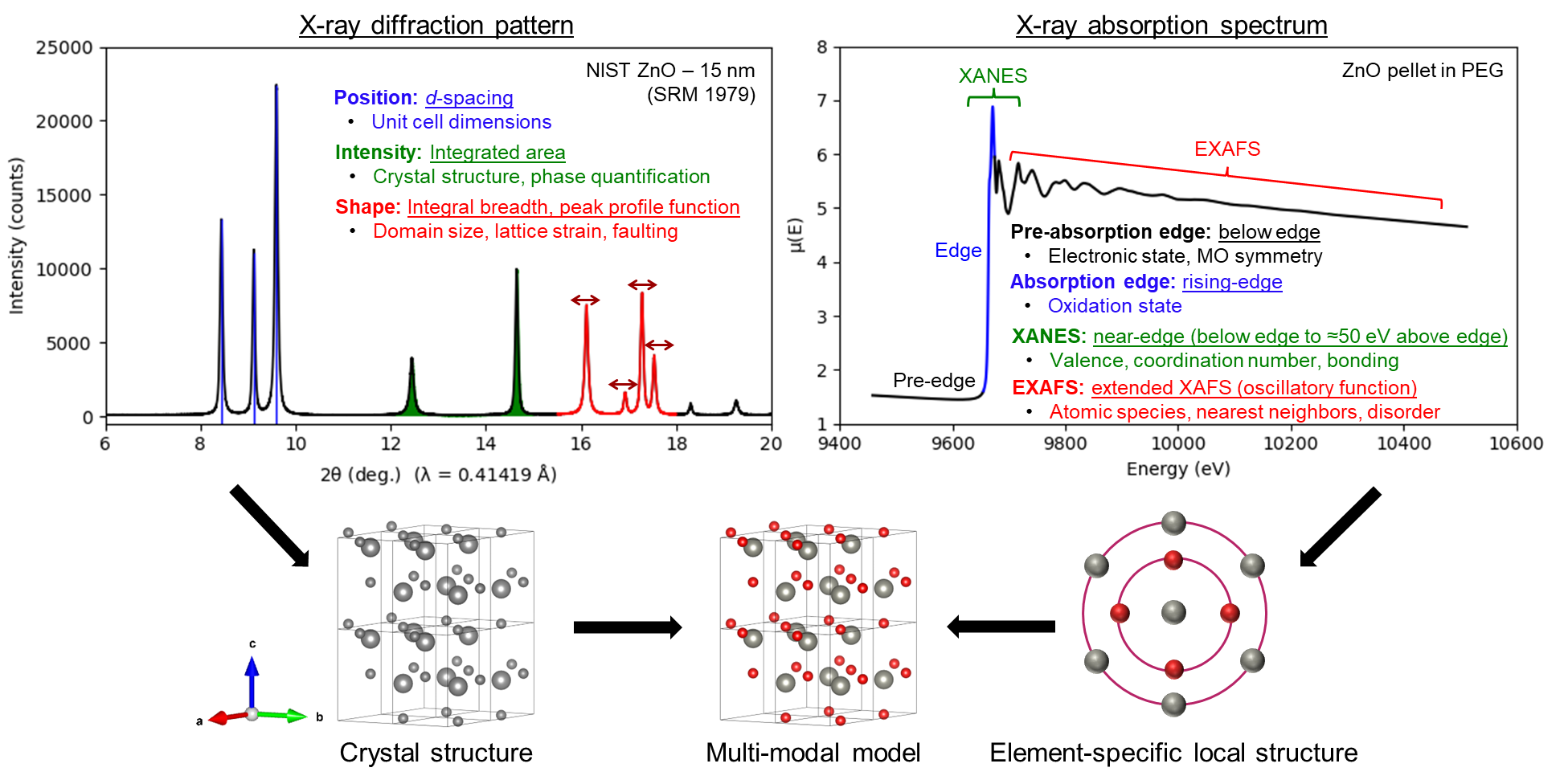}
\caption{Complementary information content of X-ray diffraction patterns and X-ray absorption spectra. In a combined analysis the partial pair distribution functions from XAFS measurements provide chemistry-specific structural constraints such as bond-distances.}
\label{fig:multimodal_data_reasoning}
\end{figure}

      Materials characterization is essential both to the discovery of new materials and for gaining an understanding of structure-property relationships. For compositionally complex and heterogeneous materials such as high entropy alloys, high entropy oxides, and materials for energy storage, multiple characterization techniques are often required to gain a holistic understanding of the material structure and chemistry (\emph{e.g.}, how atoms are arranged and what elements are present) at multiple length scales.\cite{Brahlek2022,Jiang2021,Wang2022,Sun2017} X-ray diffraction (XRD) and X-ray absorption fine structure (XAFS) spectroscopy are complementary bulk techniques routinely used for this purpose, as XRD probes the long-range (\emph{i.e.}, periodic) structure\cite{Warren1990,Klug1974,Kaduk2021} and XAFS measures short-range (\emph{i.e.}, local) structure in an element-specific manner\cite{Rehr2000,Ravel2005,Newville2014} (Figure \ref{fig:multimodal_data_reasoning}). The utility of this combined approach is well-recognized, having spurred continuous development of global modeling approaches like reverse Monte Carlo\cite{McGreevy1988,McGreevy1995,Tucker2007,Krayzman2008,Krayzman2009,Nmeth2012}, and is a powerful combination for studying complex materials including those relevant to catalysis\cite{Clausen1998,Ehrlich2011,Frenkel2011}, energy storage\cite{Sun2017,Takabayashi2024,Cao2024,Sottmann2016}, nuclear containment\cite{Fabian2021,Nicholls2022}, and cultural heritage.\cite{Pouyet2015,Cotte2018} 
      
      Researchers frequently visit large-scale user facilities such as synchrotron light sources to perform these multi-modal measurements due to their many advantages over lab scale instruments. These include enhanced spatial and temporal resolution, better sensitivity afforded by a higher photon flux, and improved compatibility with \emph{in situ} and \emph{operando} modalities. In the case of devices and samples such as battery cells and reaction vessels (\emph{e.g.}, containing catalysts), spatially resolved scattering and spectroscopic studies are needed to identify chemical and structural heterogeneities that impact reactivity, properties, and performance.\cite{Liu2016,Li2020,Decarolis2021,Meirer2018,Dann2019} For example, mapping Li-ion and Li-metal battery pouch cells with synchrotron-XRD reveals that the average state of charge inferred from voltage/current profiles is not always representative of the spatial inhomogeneity within the electrodes, and that multiple complex failure mechanisms related to inhomogeneity can be responsible for reduced lifetime and capacity.\cite{Mattei2021,Cosby2022} Inhomogeneities which are often present in dynamic chemical systems are challenging to characterize, but can reveal informative phenomena and mechanisms that would otherwise be missed.\cite{Tanim2020,Grunwaldt2005,Kelly2015,Gnzler2015} Similarly, combinatorial material libraries often synthesized as thin films by co-depositing multiple materials onto a substrate also require spatial mapping studies to understand localized variations in composition, structure, and properties. These sample libraries comprising 1,000s of individual samples are synthesized for the purpose of efficiently exploring complex composition spaces (\emph{e.g.}, ternary, quaternary, and quinary phase diagrams) with the goals of discovering new materials and gaining an understanding of synthesis-process-structure-property relationships.\cite{Kennedy1965,Xiang1995,Ludwig2008} 
      
      Regardless of whether a researcher aims to study 1000 individual samples or map a combinatorial library, facility access time is an extremely limited resource and measurement efficiency needs to be maximized. There is necessarily a limitation on the number of samples or states (\emph{e.g.}, steps in temperature, pressure, voltage, applied stress) that can be measured given that the form factor of many mappable samples and devices are on the scale of 100 mm to 500 mm while the measurement probe size (\emph{i.e.}, the X-ray beam) is on the order of 0.01 mm to 0.5 mm. For example, full-resolution mapping (defined as measuring every point available from dividing the sample area by the beam size) of a circular sample with a 60 mm diameter (a common size for wafers which combinatorial libraries are deposited onto) with X-ray diffraction would take approximately 2 days ($\approx$10 s collection time) using the Pair-Distribution Function (PDF) beamline at the National Synchrotron Light Source-II (NSLS-II) at Brookhaven National Laboratory (Figure S1). More time-consuming measurements exacerbate this issue, such that it would take approximately 4 months to collect a complementary full-resolution XAFS map for only a single elements absorption edge ($\approx$10 min.\@ collection time) using the Beamline for Materials Measurement (BMM) at NSLS-II (Figure S1). Additionally, researchers do not aim to study just a single sample, but many samples, such as those prepared under different conditions including temperature, pressure, and chemical environment. As beamtime allocations are typically on the order of a few days, conventional measurement strategies are to either perform a raster scan with a fixed step size (\emph{e.g.}, snaking line scan gridding the measurable area) or collect a geometric series that provides grids of points in which the spatial resolution increases with each subsequent grid (\emph{e.g.}, step size evolves to provide a finer grid) (Figure S2). However, these homogeneous mapping approaches are inherently mismatched with inhomogeneous samples because the scientifically valuable information (\emph{e.g.}, chemical composition, crystallographic phase distributions, atomic coordination environment) does not necessarily evolve uniformly in space. This means that the scientific understandings sought (\emph{e.g.}, phase boundaries and transitions, shape of interfaces) may not be discernible from the measurements done, resulting in missed scientific opportunities and inefficient use of beamtime which impedes both discovery and facility throughput.

      The limitations of conventional approaches to phase mapping, materials discovery, and experiment design have led to the development of data-driven, algorithmic approaches, with especially rapid development in the last decade as machine learning (ML) models and artificial intelligence (AI) have become more accessible and easily deployable.\cite{Adams2024,Chen2021,Maffettone2021b,Stanev2018,Maffettone2021b,LeBras2011,Kusne2020,Maffettone2021a,Lee2020,Long2009,Ermon2012,Hartigan1979,Pogue2023,LeBras2014,Chitturi2024,Carbone2024} This includes methods for signal processing and classification\cite{Maffettone2021b,Long2009}, Bayesian optimization-based active learning\cite{Kusne2020,Chitturi2024} and a formula to quantify the scientific value of measurements and balance exploration and exploitation,\cite{Carbone2024} though this is just a snapshot of the myriad of tools available and new ones are constantly in development. These ML- and AI-based approaches are powering materials acceleration platforms,\cite{Hse2019,Seifrid2022,Stach2021,Stier2024} also known as ``self-driving'', closed-loop, or autonomous labs, that typically have well-defined objectives such as discovering a higher temperature superconductor\cite{Pogue2023} or a novel electrocatalyst.\cite{Moon2024} These approaches can also enhance outcomes at user facilities where both the value of individual measurements to understanding a system and possible novel observations are not known \emph{a priori} (\emph{e.g.}, detect subtle phase transitions,\cite{Maffettone2021c} identify novel phases and mixtures\cite{Banko2021}).\cite{Maffettone2021a,Konstantinova2022,Maffettone2022,Konstantinova2022,Barbour2022} For example, neutron diffraction-based strain mapping can be accelerated using Bayesian optimization\cite{Snoek2012} of a Gaussian process regression\cite{Rasmussen2005} to iteratively perform measurements, infer a strain field model based on the collected data, and predict the next most important measurement, such that only 30\% to 40\% of the full dataset is needed for reconstruction.\cite{Venkatakrishnan2023} This type of active learning\cite{Settles2009} approach in which training of a ML model occurs within the measurement loop is effective for driving characterization of mappable samples using techniques such as small-angle X-ray scattering,\cite{Noack2019} angle-resolved photoemission spectroscopy,\cite{Melton2020} scanning probe microscopy,\cite{Ziatdinov2020} and X-ray diffraction.\cite{Kusne2020} 
      
      While ML and AI-driven approaches have been successfully deployed in both home labs and at user facilities\cite{Kusne2020,McDannald2022}, previous implementations typically have specific computational frameworks tailored to individual tasks rather than being generalizable. Broader needs identified by the materials discovery community include facility-wide integration, community development, extensibility towards diverse characterization techniques, direct instrumentation control, and the incorporation of AI agents.\cite{Seifrid2022,Stach2021,Stier2024,Maffettone2023b,Kusne2023,Statt2024} These criteria are paramount at large-scale user facilities that serve an interdisciplinary user base with a diversity of backgrounds and expertise. Further, the modularity of experiments done on an individual beamline as well as across beamlines requires adaptive workflows in which processes can be turned on and off with ease. ML and AI methods have great potential to enhance the efficiency of large-scale facilities but require a flexible framework that enables collaborative human-AI workflows in which users maintain autonomy over experiments and AI assists. While terminology around AI continues to evolve and varies by field, we define an AI agent in this context as a modular, task-specific component of an autonomous workflow that ingests inputs and executes bounded actions via human defined interfaces without necessarily requiring the use of large language models.

      Here we describe the development of a modular framework for AI-driven, multi-modal synchrotron experiments and the successful live orchestration of experiments using multiple beamlines with measurement times differing by several orders of magnitude. Through our decentralized approach an ensemble of AI agents performs data processing, analysis, and interpretation, leveraging all data available in real-time to guide both XRD and XAFS measurements on identical combinatorial libraries of a ternary alloy. Agents' measurement requests are moderated by human controlled adjudicators which are intermediaries between measurement requests and execution that provide an interface for researchers to query decision-making, add their own plans alongside AI agents, and control agents' roles. This ensures that human experts can evaluate agent output, engage in decision-making, and maintain autonomy. A digital twin of this framework was developed to facilitate the design and optimization of agents without consuming precious beamtime, resulting in AI-driven measurement strategies that surpass conventional approaches for phase mapping. This work was enabled by the Bluesky\cite{Allan2019,Arkilic2017} software suite as well as recently developed technologies for both secure data streaming and instrument controls,\cite{Rakitin2022}  and provides a platform for orchestrating collaborative human-AI experiments at NSLS-II and beyond.\cite{Maffettone2023a}

\section{Results}

\subsection{Architecture for multi-beamline, multi-modal AI-driven measurements.}
\begin{figure}[b]
\centering
\includegraphics[width=0.5\textwidth]{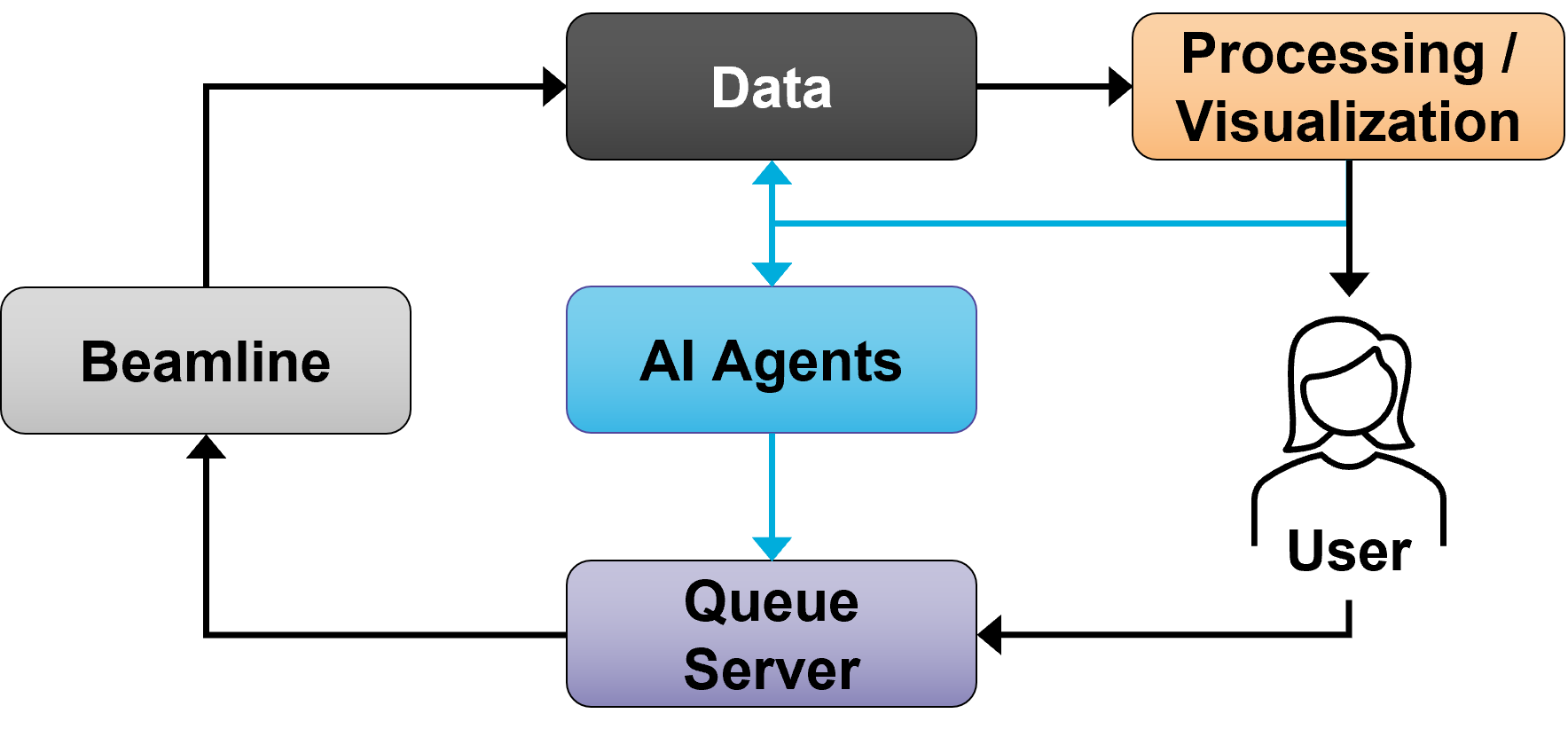}
\caption{Flow diagram of a conventional beamline architecture (outer loop) at NSLS-II and the incorporation of AI agents to the workflow. This generalized workflow is expanded upon in Figure \ref{fig:multimodal_blockdiagram} with details about agents used in autonomous experiments done at the PDF and BMM beamlines, as well as \textit{in silico}.}
\label{fig:simple_meas_loop}
\end{figure}

      At the most basic level beamline operations require a user to request a specific measurement, an interface for handling those requests, instrument controls, and data storage. At NSLS-II, one such workflow (outer loop in Figure \ref{fig:simple_meas_loop}) begins with a user requesting measurements, then these requests are added to a queue of measurement plans managed and executed by a queue server that has secure access to instrument controls on a beamline. After a measurement is complete data is stowed in a repository from which automated data processing and visualization tools make this data available to users. This workflow is enabled by existing architecture at NSLS-II for data acquisition and instrument controls (Bluesky), measurement management (queue-server\cite{Rakitin2022}), data storage and remote access (Tiled\cite{BrookhavenNationalLaboratory2025a}), and data interaction (JupyterHub\cite{Milligan2017,Rind2020}). We use the same infrastructure to incorporate AI agents as a central component that interfaces with data repositories, the measurement queue, and users to capitalize on the many opportunities for data-driven insights and guidance (Figure \ref{fig:simple_meas_loop}). A key advantage of this approach is that AI agents become a tool in the inventory of the beamline or facility rather than a core function on which operations depend. This is just one example of how AI can be integrated at a beamline that can be generalized for the scope of a large-scale facility.
      
      There are a few key needs for a robust and supportable computational framework that spans a large-scale facility. First, the framework must be generalized and extensible to other beamlines, facilities, and alternative data sources such as historical data and simulations (\emph{e.g.}, density functional theory calculations) because all of these modalities contribute to the understanding of a system that informs experimental decisions. Second, the framework must be part of an open-source project. This encourages community development efforts that extend functionality rather than duplicating the work of other projects and facilitates the capture of domain-specific knowledge and best practices in tailored agents. These first two points both address the interdisciplinary nature of synchrotron science which requires that a computational framework be flexible and accommodate domain-specific agents and workflows, as data processing, analysis and interpretation differ between techniques and fields of study. Third, the framework must enable a range of collaborative human-AI operating modes from fully human-driven to fully machine-driven with an interface for users to monitor, modify, and disable autonomous operations. Lastly, a framework that incorporates AI must provide safe, secure, and reliable operations - a requirement of any hardware or software deployed at a user facility. This last point is the most critical of all and is why in our design AI agents pass measurement requests to adjudicators (adjudication is discussed in the following section) that then assemble measurement plans and send these to a secure queue server that has allowable measurements defined by humans. Execution of these requests depends on the state of the safety interlock system which agents cannot interact with. Further, agents do not have knowledge of the application programming interface (API) for the interlock system.

\subsection{AI agents for data processing, analysis, and decision-making.}
      With the requirements for a flexible, open, and reliable computational framework in mind we designed a modular suite of AI agents to perform common tasks associated with XRD and XAFS measurements. We also developed a facility-controlled adjudicator to plug agents into that only passes allowable instrument control requests to the queue server. Agents are built with plug-and-play capabilities to ensure they can be easily integrated, extended, or replaced as experimental needs evolve. Accessible configuration controls allow users to modify agents, such as their parameterization for automated analysis or role in decision-making for an experiment. Configuration can be hard coded into the agent design, supplied upon initialization, or modified through a graphical user interface (GUI) depending on the agent's intended use. Another important aspect of our agents is that their outputs can be queried through Tiled in the same manner that data is accessed, taking advantage of the existing digital infrastructure to expose agent behavior and decision-making. Lastly, agents are deployed on virtual machines, enabling scalable and distributed computing which is needed for several of the computationally demanding processes that facilitate autonomous experiments.
      
\begin{figure}
\centering
\includegraphics[width=\textwidth]{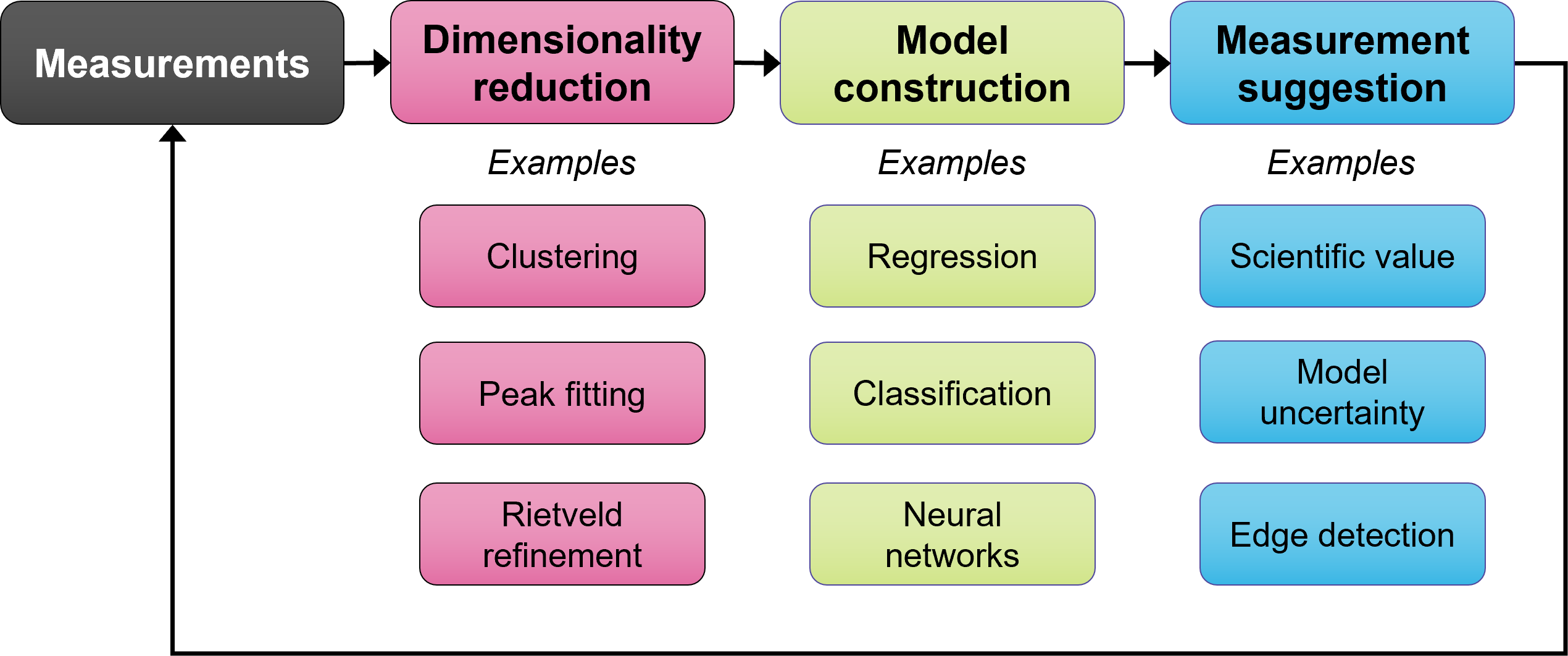}
\caption{Bayesian optimization-based autonomous experiment workflow with examples of AI agents for data dimensionality reduction, model construction, and measurement suggestion. Tasks can be performed by many different AI agents (sublevels) individually or in concert, with agent outputs as input to subse-quent agents. This workflow is used for the decision-making layer (Figure \ref{fig:multimodal_blockdiagram}) that drives autonomous experiments at the PDF and BMM beamlines, as well as \textit{in silico}, and is highly extensible.}
\label{fig:BO_loop}
\end{figure}

      Autonomous operating modes are enabled by connecting the inputs and outputs of our separable agents in a Bayesian optimization (BO) process (Figure \ref{fig:BO_loop}).\cite{Snoek2012} BO is an iterative method for optimizing a black-box function that has become a cornerstone in autonomous experimentation.\cite{Adams2024,Stanev2018,Kusne2020,Strieth-Kalthoff2024,Pogue2023,Chitturi2024,Hse2019,Seifrid2022,Stach2021,Moon2024,Maffettone2023a,Venkatakrishnan2023,Noack2019,McDannald2022,Maffettone2023b,Kusne2023} The application of BO to self-driving phase mapping can be best understood by first generalizing these studies as presenting an experiment design question ``how do we explore the unknown?'' This problem can then be broken down into discrete tasks: (1) establish a basis of understanding through initial observations, (2) develop a model from observations, (3) make further observations based on insight from the model, and (4) refine the model based on data.\cite{Galilei1914} While this active learning scheme is abstract, it is fundamental to the way humans and machines process information\cite{Lindsay2013,Simon1964} as well as to the scientific method itself.\cite{Galilei1914,Gauch2003} Further, the individual tasks are programmatic and well-suited to be automated by artificial intelligence. We implemented agents for data dimensionality reduction, model construction, and measurement suggestion, with both developed agents and further examples of methods that can be applied to these tasks outlined in Figure \ref{fig:BO_loop}. We broadly characterize agents as passive or active and distinguish them by their ability to suggest measurements. Passive agents can query measured data, perform analyses, and report results, while active agents have the additional ability to suggest measurements.

      Our decision-making routine begins with data dimensionality reduction - a process also known as feature engineering that is often a critical step in preparing data for input to ML models - by K-means clustering. K-means is a clustering algorithm that groups data into N clusters by repeatedly calculating the Euclidean distance between each data and the average location of data within the cluster (\emph{i.e.}, the centroid or cluster center) and assigning each data to the cluster whose centroid is closest until the clusters no longer change.\cite{Steinhaus1956,Lloyd1957,Lloyd1982,Ball1965,Forgy1965,MacQueen1967,Hartigan1979} This reduces each 1D diffraction pattern from a vector with thousands of elements to an integer label, and establishes a basis for understanding the data such as by visualizing the spatial distribution of clusters. While traditional data analysis methods like peak fitting and Rietveld refinement\cite{Rietveld1969} also provide reduced data by extracting what can be more direct and interpretable scientific insights such as unit cell parameters, the understanding obtained is necessarily biased by the chosen model (\emph{e.g.}, crystallographic phase, refinement parameterization). In our testing, automated Rietveld is not reliable due to the evolution of multiple symmetry-related crystalline phases that have significantly overlapped peaks. These features introduce ambiguity to the refinement parameterization and extracted peak intensities (\emph{i.e.}, integrated area) so we currently opt for a more reliable dimensionality reduction method with significantly fewer parameters as part of the active learning process. Fortunately, the plug-and-play aspect of our modular framework allows us to swap it out with a better tool when it becomes available or enable it as an active agent when its performance is more reliable such as when exploring a known phase space for which refinements can be appropriately parameterized.
      
      After the dimensionality of the data is reduced, the next step is to develop a model based on observations. One may be inclined to reach for continuous functions such as linear interpolators or cubic splines that are commonly used for modeling low-dimensional data with minimal noise, but these are ill-suited to higher-dimensional data for which an appropriate model is often impractical to derive and optimize, especially when considering that these data often have hidden variables. In these instances, machine learning models such as a Gaussian process (GP) regressor can be more appropriate as they compute a probabilistic model that approximates the objective function describing the data by optimizing the mean function and its covariance (defined by the kernel selected) based on input data.\cite{Rasmussen2005} Generally, a GP is trained on some observables such as 1D data and the corresponding x,y coordinates, then a surrogate model of the dataset is obtained by predicting the observables at coordinates where measurements have not been done. Instead of training a GP on 1D data or integer labels derived from clustering, we compute a scientific value function\cite{Carbone2024} (SVF) from the data that describes the scientific value of each measurement for understanding the complete dataset. The scientific value for all possible measurements is then predicted and provides the surrogate model that is needed as part of the measurement selection step in BO. The flexibility of our framework streamlines the future development of alternative agents for model construction such as Bayesian neural networks\cite{Jospin2022} that integrate uncertainty quantification or ensemble methods\cite{Dietterich2000} like random forests\cite{Breiman2001} that can capture complex non-linear spatial relationships.
      
      The final step in our autonomous measurement loop is to select the next measurement(s) based on the potential information gain to the surrogate model. Information gain is assessed through an acquisition function derived from the model. Acquisition functions commonly used in Bayesian optimization include expected improvement\cite{Mokus1975}, upper confidence bound\cite{Srinivas2012}, and uncertainty quantified via Shannon entropy\cite{Shannon1948,Hennig2012}, each of which differently balances exploration of regions with sparse measurements and exploitation of regions with high predicted value. Once an acquisition function is computed the next measurement is often selected from the maximum of the function such as the position where scientific value or model uncertainty is highest. Alternatively, measurements can be sampled from a probability distribution derived from the acquisition function to avoid highly localized measurements. After the selected measurement is done, the autonomous loop continues with the inclusion of the newly acquired data. 

\subsection{Multi-modal experiment at PDF and BMM.}
\begin{figure}
\centering
\includegraphics[width=\textwidth]{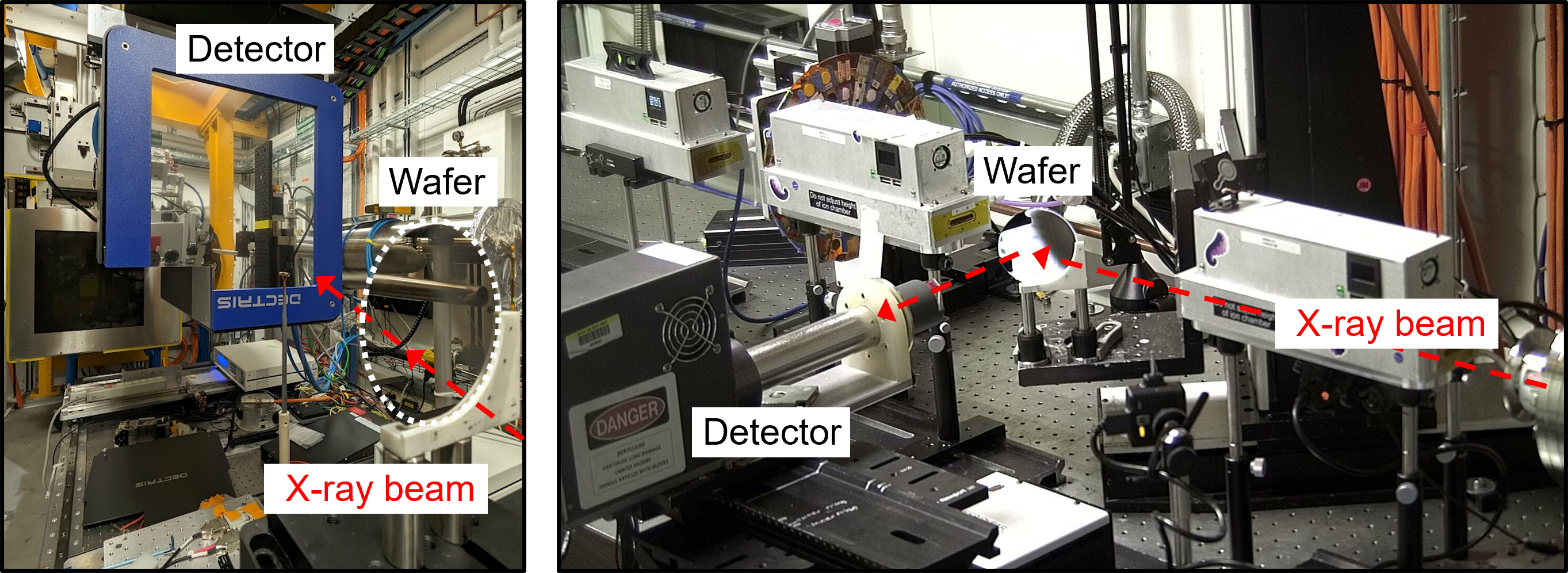}
\caption{Al-Ni-Pt thin films prepared on an amorphous glass substrate mounted on the PDF beamline for XRD measurements (left) and BMM beamline for XAFS measurements (right).}
\label{fig:samples_mounted}
\end{figure}

      Synchronous mapping of a combinatorial library was orchestrated on the PDF and BMM beamlines at NSLS-II to demonstrate that our framework and suite of agents enable multi-modal, multi-beamline experiments. While complementary diffraction and spectroscopy studies are typically conducted asynchronously, with faster XRD measurements done first, then slower spectroscopy measurements done separately in a later beamtime, it was important that we demonstrate that synchronous modes are possible. To test this, we prepared two identical Al-Ni-Pt thin film combinatorial libraries by sputter deposition on an amorphous glass substrate (see Methods for synthesis details). This system was selected because it met our experimental requirement for samples with a diversity of features and strong scattering signals that could be reproduced. A coarse grid of XRD measurements was done on each film to confirm that they were qualitatively identical. These ternary alloy combinatorial libraries represent a compositional slice of the Al-Ni-Pt phase diagram such that the relative amounts of Al, Ni, and Pt, crystalline phases, and local coordination environment evolve as a function of position. Each wafer was mounted in a 3D-printed holder that minimally obscures the sample on the PDF and BMM beamlines (Figure \ref{fig:samples_mounted}). Measurements were aligned to a common coordinate system with the wafer center as the origin by finding 3 points along the circumference of the wafer for triangulation.

\begin{figure}
\centering
\includegraphics[width=\textwidth]{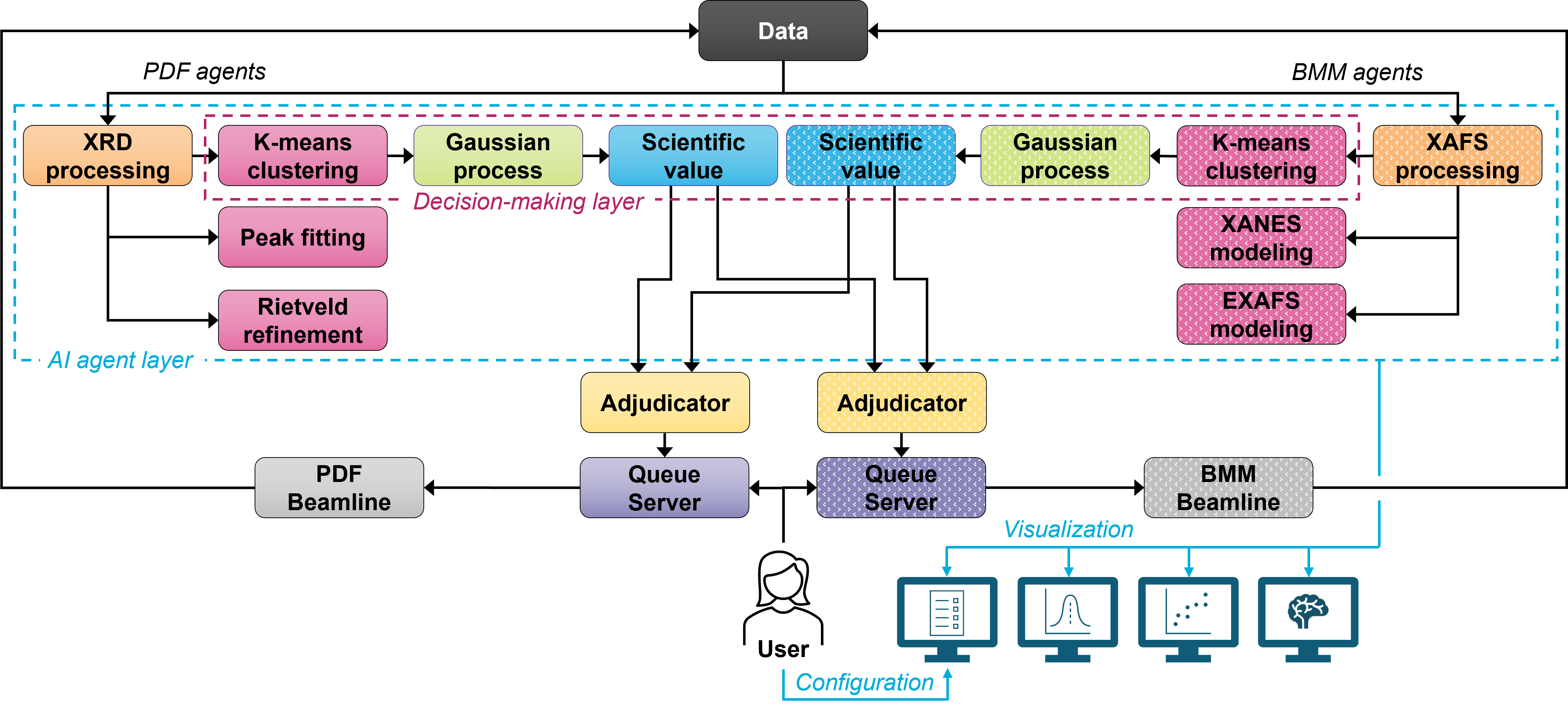}
\caption{Block diagram of software used in multi-modal AI-driven experiments with human-in-the-loop to orchestrate synchronous XRD and XAFS measurements. All agent outputs are stored and visualized so human experts can evaluate decision-making as well as results from analysis including XRD and EXAFS modeling. This workflow is generalized in Figure \ref{fig:simple_meas_loop}, and further examples of agents for the decision-making layer are provided in Figure \ref{fig:BO_loop}}
\label{fig:multimodal_blockdiagram}
\end{figure}

      Automated data processing, reduction, analysis, and decision-making were needed for both the XRD and XAFS measurements. A diagram of the multi-beamline computational workflow used for these experiments is provided in Figure \ref{fig:multimodal_blockdiagram}, expanding on the generalized representation in Figure \ref{fig:BO_loop}. The main features are remote data storage, an AI agent layer (dashed blue outline) for data processing, analysis, and decision-making (dashed pink outline), adjudicators for each beamline that handle measurement requests from decision-making agents, queue servers that prove a common access point for humans and agents to add measurements, and GUIs for visualization and agent configuration. AI agents specific to the PDF beamline are on the left side of the diagram in Figure \ref{fig:multimodal_blockdiagram}, while those for BMM are on the right side. Agent outputs are stored and available for visualization whether they are active in decision-making or passively performing analysis as is the case for XRD and XAS modeling agents.
      
      Initial data collected from seeding measurements with either random sampling or a small coarse grid is queried by data processing agents for each beamline (orange boxes). For diffraction measurements this includes azimuthal integration of 2D images to 1D patterns, intensity normalization, and background subtraction. XAFS data processing involves multiple steps, including: (1) normalizing absorption spectra ($\mu$(E) \emph{vs.} E) to yield X-ray absorption near-edge spectroscopy (XANES) data; (2) background subtraction which provides extended X-ray absorption fine structure (EXAFS) data as $\chi$(E) \emph{vs.} E; (3) conversion from energy (E) to wavenumber (k) space (\emph{i.e.}, conventional EXAFS in the form of $\chi$(k) \emph{vs.} k; and (4) the forward Fourier transform from k-space to R-space (|$\chi$(R)| \emph{vs.} R) which is the form of EXAFS data suitable for modeling to extract partial pair distribution functions.\cite{Kelly2015} 
      
      Processed data is then streamed to the data analysis and dimensionality reduction agents (pink boxes) that can provide more detailed scientific insights and prepare data for use in the decision-making layer. For example, our diffraction peak fitting agent outputs peak intensities, positions, and widths, parameters that are systematically impacted by the structural, chemical, and morphological changes in a sample that can be difficult to discern directly from 1D data. An automated Rietveld refinement agent was also deployed using the python scripting interface to GSAS-II,\cite{Toby2013,ODonnell2018} with refinement parameterization provided by the user. Detailed output from the refinement agent is stored for visualization and expert feedback, including heuristics such as R-factors and fit quality, crystallographic details like unit cell parameters and cell volume, and phase fractions when multiple phases are present. On the XAFS-side, automated XANES and EXAFS modeling agents use Larch\cite{Newville2013} to extract details such as the number of nearest neighbors to an absorbing atom in a chemistry-specific manner.
      
      Autonomous operations were enabled by agents in the Bayesian optimization driven decision-making layer (Figure \ref{fig:multimodal_blockdiagram}, dashed pink outline). First, data dimensionality reduction was done by K-means clustering on processed diffraction patterns and XAFS spectra ($\mu$(E) \emph{vs.} E). Next, a scientific value function\cite{Carbone2024} was calculated from the measurement positions and clustering labels, and a surrogate model was constructed by training a GP regressor with a Matern kernel on the scientific value function. Lastly, the upper confidence bound of the scientific value function was used as an acquisition function to select the next measurement based on the location with the highest predicted scientific value. Here, the scientific value functions for XRD and XAFS are independent of one another. Though decision-making agents were deployed for both the XRD and XAFS data streams, the massive difference in measurement time ($\approx$10 s \emph{vs.} $\approx$10 min.\@) led us to enable only the XRD-based measurement suggestions for active learning and configure the XAFS-based agent to report but not add to the queue. In this operating mode, measurement suggestions from the PDF beamline were added to the queue servers on both the PDF and BMM beamlines, with adjudicators in between to eliminate redundancies.
      
      In this experiment we achieved inter-beamline measurement requests, inter-agent communications, on-the-fly data processing, analysis, and interpretation, and autonomous operations. Several algorithms for autonomous exploration were tested and a conventional grid was also collected. During the allotted beamtime it was discovered that decisions made by those versions of the agents were too heavily prioritizing exploitation over exploration, leading to many localized measurements near the wafer edge and sparse measurements in regions that were not sufficiently explored. This demonstrated the need for a range of operating modes with varying degrees of autonomy as well as agent configurations that can be easily modified.

\subsection{Post-experiment analysis of measured data from Al-Ni-Pt thin films.}
      XAFS and XRD data were analyzed to assess the composition and structure as a function of position on the thin film. This includes XAFS data that was collected for both the Pt \textit{$L_3$} and Ni \textit{K} edges on a grid of 49 positions spaced 8 mm apart, and XRD measurements that were done approximately 1.2 mm apart providing a higher resolution map consisting of 1876 points. Due to the sensitivity of XAFS to the local chemical environment about an absorbing atom that is independent of the long-range ordering or lack thereof (\emph{i.e.}, applicable to crystalline and amorphous materials, liquids, and gases alike), composition can be determined from these data within the inherent limits and uncertainties of the measurement and modeling approach used.\cite{Newville1999,Chantler2024} While extracting composition from diffraction data is challenging, especially when multiple phases with variable composition are present, the crystalline phases present and relative changes in cell volume can be determined.
      
\begin{figure}
\centering
\includegraphics[width=\textwidth]{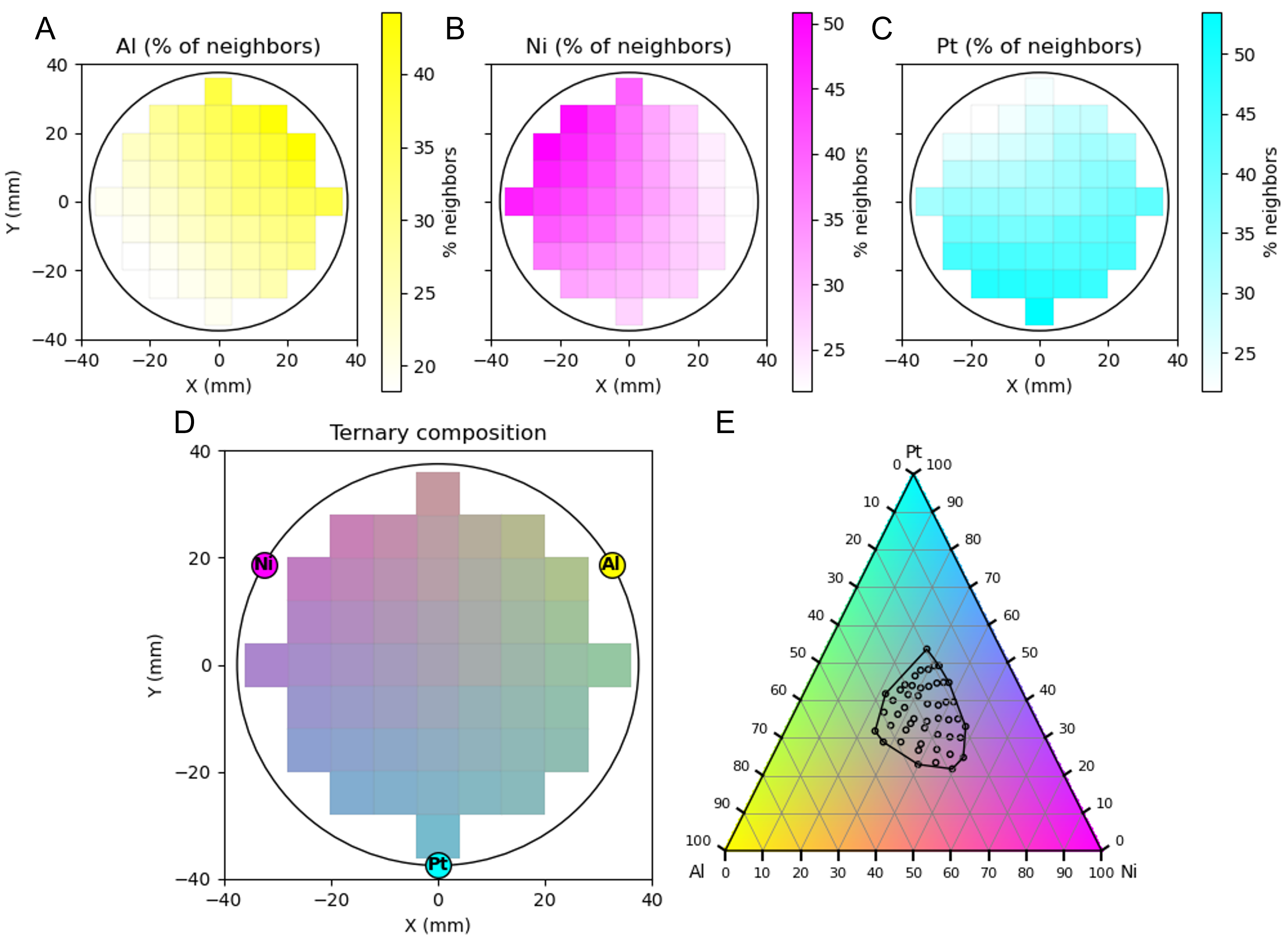}
\caption{(A-C) Percentage of Pt-M (M = Al, Ni, Pt) nearest neighbors determined from modeling EXAFS data collected at the Pt $L_3$ absorption edge. (D) Composition map with approximate sputter deposition sites marked along wafer edge (black circle) and (E) ternary color legend with markers for compositions extracted from XAFS measurements and a polygon bounding the composition range studied.}
\label{fig:EXAFS_results}
\end{figure}

      Reduced EXAFS spectra from the Pt L3 edge were modeled assuming a face centered cubic coordination environment (12 nearest neighbors) to extract the number of Al, Ni, and Pt nearest neighbors (Figure S3), with an example of the fit quality shown in Figure S4. Details of the EXAFS analysis are explained in the SI. Relative amounts of Al, Ni, and Pt (Figures \ref{fig:EXAFS_results}A-C) can be determined from the number of nearest neighbors, and composition as a function of position can then be visualized using a ternary color map (Figure \ref{fig:EXAFS_results}D-E). This demonstrates that the abundance of each metal is highest at positions closest to its deposition site (marked in Figure \ref{fig:EXAFS_results}D) and decreases smoothly with increasing distance from the deposition site. Our analysis found that Al, Ni, and Pt are present in the ranges of 18.2 \% to 44.3 \%, 21.8 \% to 50.9 \%, and 21.8 \% to 53.6 \%, respectively, such that there are no regions where the abundance of an individual element exceeds 53.6 \% or is less than 18.2 \%. This behavior can be more easily understood from a ternary diagram (Figure \ref{fig:EXAFS_results}E) in which the open circles mark the discrete compositions determined by fitting EXAFS spectra and the bounding polygon encapsulates the compositional space studied.

      While the peaks present in the XRD data vary significantly as a function of position on the thin film, with some peaks disappearing and appearing indicating multiple evolving phases, both a face-centered cubic (FCC, \textit{Fm$\bar{3}$m}, space group \#225) and a face-centered tetragonal (FCT, \textit{P4/mmm}, space group \#123) phase are found (Figure S5). Due to the sub-cell relationship that the FCC phase shares with the FCT phase, many of the peaks are highly overlapped, including the most intense peak at approximately $4.4^\circ$ ($2\theta$). Additionally, the relative intensities of the low angle FCT peaks ($<4^\circ$) vary significantly with respect to the major peak and have considerable uncertainty due to phase-specific texturing identified in the 2D diffraction images (Figures S10-S12). These features make it challenging to distinguish when one or both phases are present from only a visual inspection of the data, though overlays of patterns collected from the bottom of the film to the top (Figure S5A) and from left to right (Figure S5B) suggest that only data collected in the upper left region of the film can be described solely by the FCC model.
      
\begin{figure}[t]
\centering
\includegraphics[width=\textwidth]{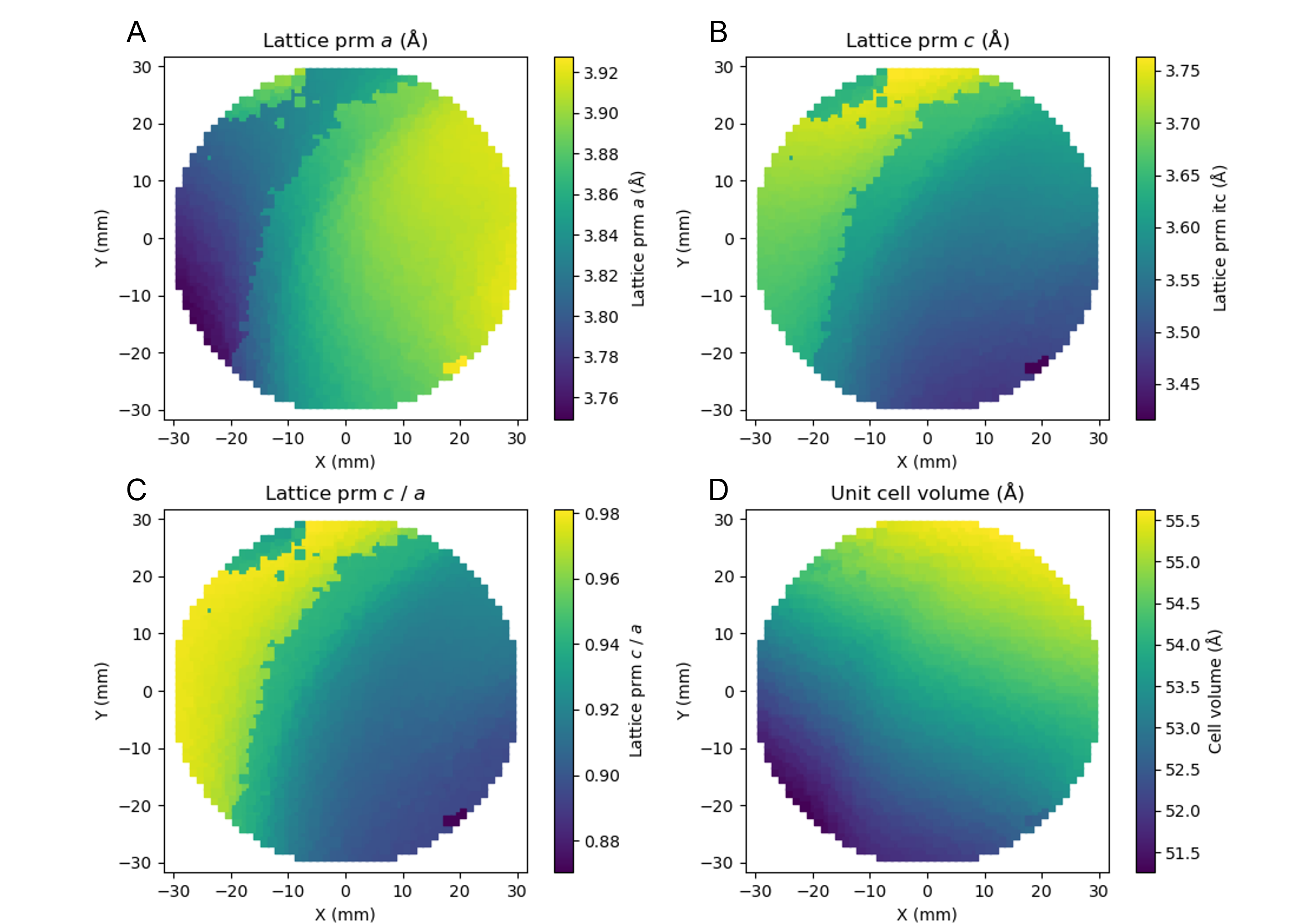}
\caption{X-ray diffraction ($\lambda$ = 0.1665 \AA) mapping results. (A-D) Unit cell details extracted by Pawley refinement using a tetragonal (space group \#123, \textit{P4/mmm}) solid solution phase. (A) Unit cell lattice parameter \textit{a} (\AA). (B) Unit cell lattice parameter \textit{c} (\AA). (C) Lattice parameter \textit{c} / \textit{a} ratio. (D) Unit cell volume \AA$^3$.}
\label{fig:AlNiPt_XRD_results}
\end{figure}

      Whole pattern fitting of the azimuthally integrated 1D patterns was done via Rietveld\cite{Rietveld1969} and Pawley\cite{Pawley1981} methods to compare the validity of models across the film and extract changes in peak positions that are a result of compositional changes from alloying. Pawley fits using an FCC model yielded lattice parameters ranging from 3.70 \AA\, to 3.82 \AA\, as a function of position on the film (Figure S7A), while Al, Ni, and Pt (all of which are FCC metals) have lattice parameters of 4.05 \AA, 3.52 \AA, and 3.92 \AA, respectively. This suggests the FCC phase is a solid solution, and a map of cell volume (Figure S7B) shows that the FCC unit cell is largest at the top of the film near the Al deposition site and smallest on the bottom right side of the film. While cell volume is expected to be maximized near the Al deposition site, it is puzzling that the smallest cell volumes are found on the bottom right portion of the film because in principle the bottom left region furthest from the Al deposition should have the smallest volume. However, poor fits to data outside the upper left region of the film (Figure S7D-E) indicate the FCC model is only valid for the Ni-rich region where FCT peaks are not found. Pawley fits with the FCT phase are found to better model the data (Figure S8) and demonstrate the expected trend in cell volume compared to the FCC fits (Figure \ref{fig:AlNiPt_XRD_results}D) while also providing additional insight from the separate a and c lattice parameters. From prior studies of phase equilibria in the Al-Ni-Pt system\cite{Lu2009,Kamm1994,Meininger2003} it was found that the lattice parameters of the FCT phase and the \emph{c/a} ratio can be used to approximate composition, such that when \emph{c/a} is 0.89 to 0.92 the composition is \ce{AlNiPt2} and when \emph{c/a} is 1.0 to 1.02 the composition is \ce{AlNi2Pt}. Maps of the lattice parameters and \emph{c/a} ratio (Figure \ref{fig:AlNiPt_XRD_results}A-C) show a distinct region on the left side of the sample that is Ni-rich based on the \emph{c/a} value of 0.98. This Ni-rich region is highly localized, and there is an abrupt increase in \emph{a} and abrupt decrease in \emph{c} resulting in a sharp transition from \emph{c/a} values of $\approx$0.97 to $\approx$0.98 to $\approx$0.92. This suggests the FCT phase is Pt rich in all areas except the upper left region near the Ni deposition site. The trends in cell volume and approximate composition from the FCT lattice parameters indicate that both the FCC and FCT phases are sensitive to changes from alloying in regions where the models are valid. This behavior follows expectation based on both the relative atomic radii of the metals (Al 1.43 \AA\, > Pt 1.38 \AA\, > Ni 1.24 \AA) and the composition determined by EXAFS modeling.

      However, due to many complexities in the diffraction data, it is beyond the scope of this study to gain further structural insights such as the exact phases present, the composition of these phases, and their relative amounts. The complexities observed include phase-specific texturing in the 2D images, presence of multiple symmetry-related crystallographic phases (\emph{e.g.}, peaks can be exactly overlapped), and anisotropic peak broadening (\emph{i.e.}, broadening that is \textit{hkl}-dependent and not exclusively angle-dependent) (Figure S5 - S12). While methods exist for modeling some of these features, the concomitant effects introduce significant uncertainty in choosing an appropriate model, parameterizing refinements, and interpreting results. There are also large discrepancies in the literature for both the various binary mixtures (\emph{e.g.}, Al-Ni, Ni-Pt) and the ternary system regarding the compositional ranges over which many phases exist, what the structure of those phases are, and the phase diagram behavior in general.\cite{Lu2009,Kamm1994,Meininger2003,Hayashi2005,Gleeson2009,Grushko2012,Grushko2014,Popov2022,Cao2021} Further, many prior studies investigated phase equilibria at temperatures above $900^\circ$ C, thus sample preparation was done at high temperatures and included days to weeks of annealing time to achieve equilibrium and homogenization. In contrast, it is implausible that our samples are at equilibrium due to the rapid cooling rates up to $10^9$ K/s in sputter deposition.\cite{Bordeenithikasem2017,Kube2019} An extended discussion of the diffraction data including examples of the complexities encountered is provided in the Supporting Information.

\subsection{Digital twin for spatially resolved diffraction experiments.}
\begin{figure}
\centering
\includegraphics[width=0.5\textwidth]{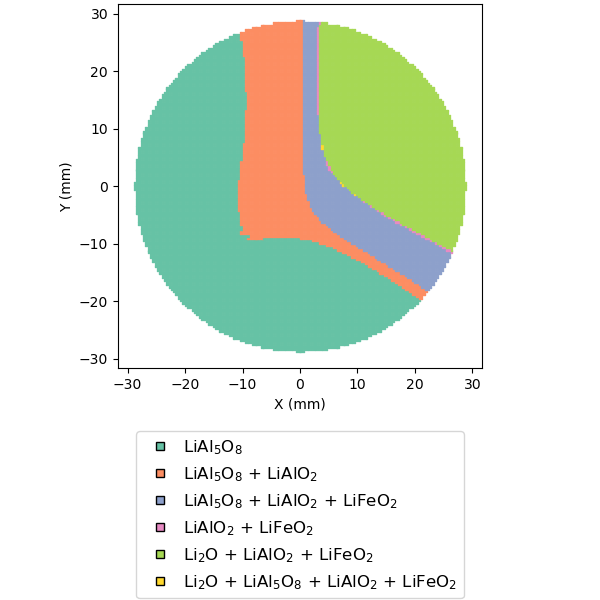}
\caption{Simulated Al-Li-Fe oxide combinatorial library with each unique combination of phases represented by a color as follows: \ce{LiAl5O8} (teal), \ce{LiAl5O8} + \ce{LiAlO2} (orange), \ce{LiAl5O8} + \ce{LiAlO2} + \ce{LiFeO2} (purple), \ce{LiAlO2} + \ce{LiFeO2} (pink), \ce{Li2O} + \ce{LiAlO2} + \ce{LiFeO2} (lime green), and \ce{Li2O} + \ce{LiAl5O8} + \ce{LiAlO2} + \ce{LiFeO2} (yellow). Corresponding distributions of each element and phase provided in Figure S13.}
\label{fig:sim_ground_truth}
\end{figure}

      We developed a digital twin of our real-world computational infrastructure for diffraction measurements to enable playback of experiments, agent development, and benchmarking of autonomous approaches (see Methods and Supporting Information). At a high level the only requirements are a dataset with coordinates, corresponding diffraction patterns, and methods for ``collecting'' data, though benchmarking requires that the ground truth phases and relative weights are known. A dataset was needed with similar features to those of our Al-Ni-Pt thin film, but the experimental considerations for a strongly scattering sample were no longer constraints. For these reasons, we turned to the literature and borrowed from a simulated diffraction dataset of the Al-Li-Fe oxide system that was designed as a testbed for autonomous experimentation\cite{LeBras2011} and has been used to validate AI-driven phase mapping.\cite{Chen2021,Stanev2018} Rather than directly sampling the phase diagram in an elemental composition space, we simulated a combinatorial library of this ternary system that is representative of both the compositional gradient and phase distributions expected on a thin film prepared by sputter deposition (Figure \ref{fig:sim_ground_truth}, Figure S13-S14).
      
      First, a mesh grid of coordinates was created for a 60 mm x 60 mm area with 0.4 mm spacing resulting in $22,500$ points. Then, the elemental composition at simulated deposition sites equidistant from the origin was defined by relative amounts of Al, Li, and Fe. Elemental composition at every point was then calculated using a Euclidean distance-based Gaussian smoothing function. Phases present, phase weights, and the resulting diffraction patterns were then interpolated based on composition from the publicly available dataset that has established relationships between these features (Figure S13-S14). Finally, points outside a 30 mm radius from the origin were excluded leaving $16,084$ points on a simulated circular sample.
      
      Across the sample each element's concentration varies from 20 \% to 60 \%, with the highest concentrations at the simulated deposition sites along the sample edge and a nearly equiatomic mixture about the center (Figure S13). Four different phases can be found in the dataset, namely \ce{LiAl5O8} (space group \#212, \textit{P\(4_{3}\)32}), \ce{LiAlO2} (space group \#166, \textit{R$\bar{3}$m}), \ce{LiFeO2} (space group \#166, \textit{R$\bar{3}$m}), and \ce{Li2O} (space group \#225, \textit{Fm$\bar{3}$m}), with maps of their relative abundance provided in Figure S13. Alloying within the phases was accounted for by relative changes in both the diffraction peak intensities and positions. The presence of a phase at each point was defined by its weight exceeding a 1 \% threshold. Then, the unique combinations of phases present across the sample were determined, resulting in 6 integer labels that serve as ground truth class labels. The spatial distribution of these unique phase combinations and the colors corresponding to the integer labels are shown in Figure \ref{fig:sim_ground_truth} (\emph{e.g.}, orange = \ce{LiAl5O8} + \ce{LiAlO2}).
      
\begin{figure}
\centering
\includegraphics[width=\textwidth]{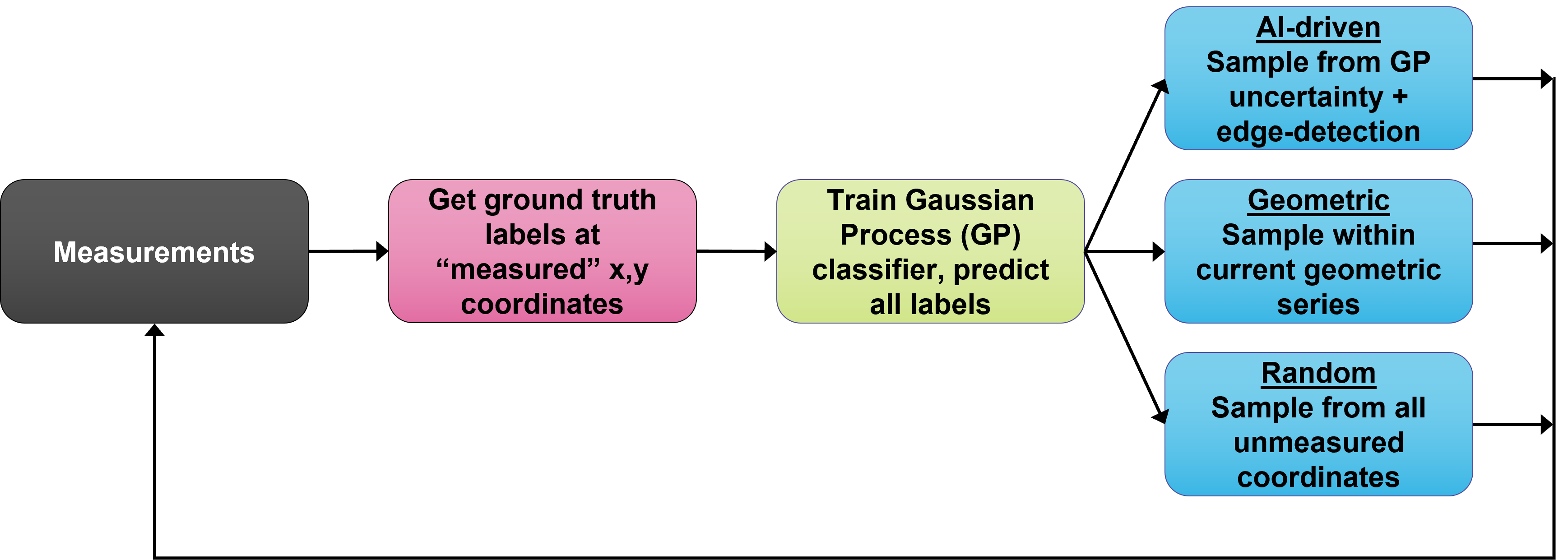}
\caption{\textit{In silico} measurement campaign logic to isolate and compare measurement three measurement selection strategies: AI-driven, geometric series (conventional grid), and random. Geometric series involves sampling from unmeasured coordinates within the current grid resolution such that the order of measurements for a given \emph{x,y} step size is randomized.}
\label{fig:sim_meas_strategies}
\end{figure}

      AI-driven phase mapping was compared to a conventional geometric series (\emph{i.e.}, grid with evolving resolution) and random sampling using a Bayesian optimization strategy (Figure \ref{fig:sim_meas_strategies}) that replaces the data dimensionality reduction by K-means clustering with retrieval of ground truth integer labels for the measured points. This removes the potential for mislabeling to interfere with the model construction and measurement selections steps. While Bayesian optimization does not drive measurement selection for the geometric series or random sampling, the model construction step (\emph{i.e.}, training and prediction using a Gaussian process classifier) is needed to compare the accuracy of models given a set of measurements. Here, accuracy is calculated by comparing the model predicted from a set of measurements (N) to the best possible model (eq. 1). 

\begin{equation}
    \text{Accuracy} = \frac{ \left(\text{Mode}l(N) + \text{Model}(N_{tot-N})\right) }{\text{Model}(N_{tot})}
\end{equation}

The best possible model (Figure S15B, denominator in eq. 1) is obtained by training a Gaussian process classifier on all x,y coordinates and corresponding ground truth labels, then predicting labels for all coordinates. The labels predicted from a set of measurements are compared to those from the best possible model rather than the ground truth because a model can only ever be as accurate as the model predicted from all the training data.

\begin{figure}
\centering
\includegraphics[width=\textwidth]{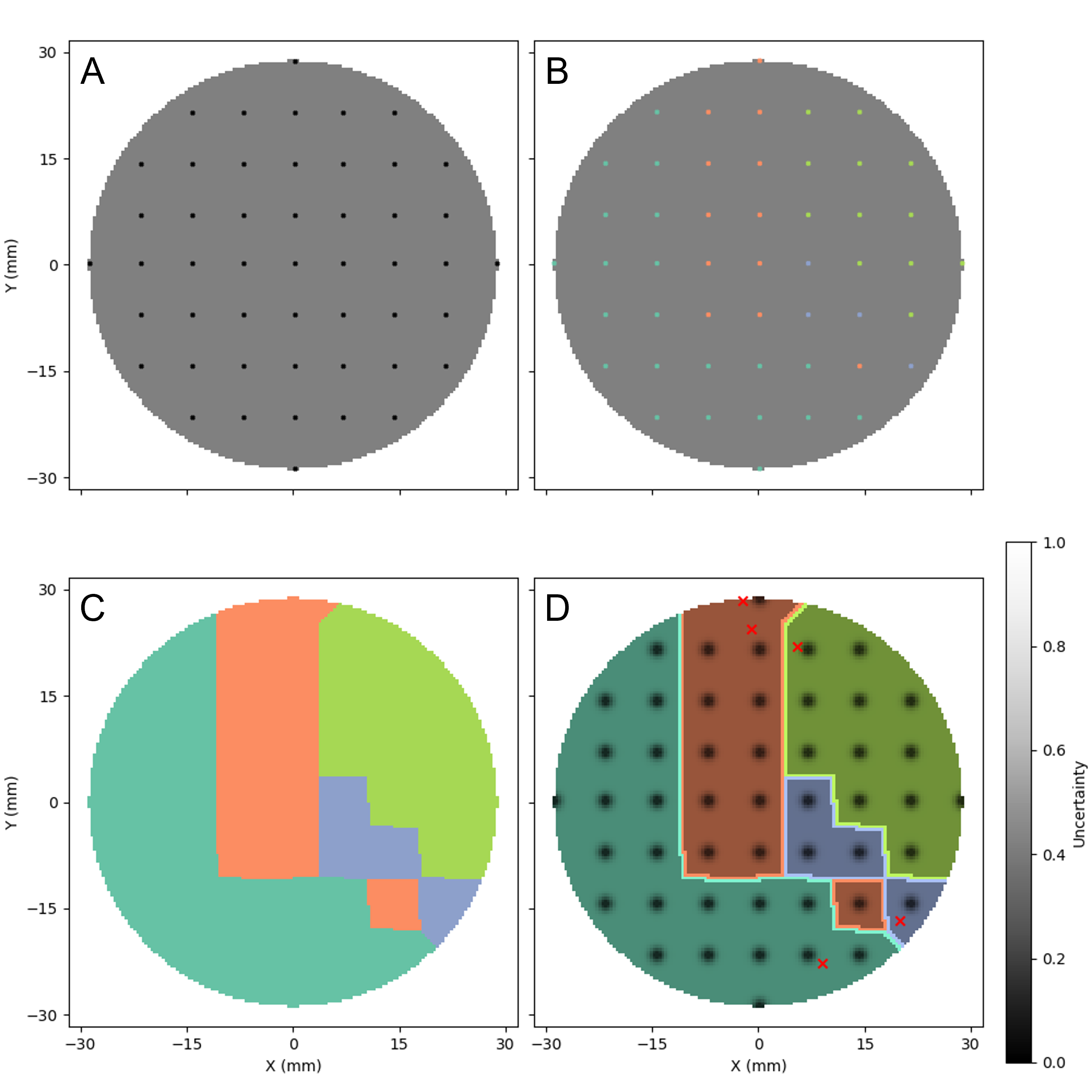}
\caption{Example AI-driven measurement strategy. A) Initial measurements on a simulated sample. B) Dimensionality reduction of 1D diffraction data to integer labels where each label is represented by a different color. C) Model of sample constructed from the labels predicted by a Gaussian process classifier trained on the measured coordinates and corresponding labels. D) Model with uncertainty from the Gaussian process classifier predictions overlaid (dark to light) and 5 measurement suggestions (red Xs) sampled from a cumulative distribution function constructed from the uncertainty.}
\label{fig:AIdriven_example}
\end{figure}

      Through this approach the accuracy of a model can be calculated after each measurement, allowing us to evaluate the performance of each mapping strategy. An example of this process is provided in Figure \ref{fig:AIdriven_example}, showing A) seeding with a coarse grid of measurements, B) labeling, and C) the predicted model. Accuracy is then calculated by comparing the predicted model to the best possible model (Figure S15B). After this step the measurement strategies diverge, with the geometric series randomly selecting an unmeasured point within the current spatial resolution (\emph{e.g.}, within the same color in Figure S2C-F) and the random sampling approach randomly selecting from all unmeasured points. For the AI-driven approach, an acquisition function is constructed by summing the normalized uncertainty associated with the classifier-derived model and a binary on/off edge characteristic (see SI for details). Rather than selecting the maximum of this function which can lead to oversampling (\emph{e.g.}, from poor minimization or a spatially localized global maxima), a probability distribution function is calculated from the acquisition function and measurements are sampled from the distribution. An example of how the acquisition function spatially varies (dark to light) and five measurement locations sampled from the probability distribution (red Xs) demonstrate how exploration of unknown regions is balanced with exploitation of locations that have high uncertainty as well as those along borders (Figure \ref{fig:AIdriven_example}D).

\subsection{Simulated AI-driven and conventional mapping campaigns.}
\begin{figure}
\centering
\includegraphics[width=0.5\textwidth]{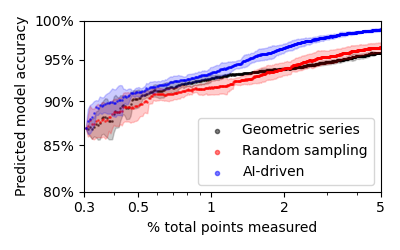}
\caption{Accuracy evolution from averaging 10 simulated experiment campaigns in which the measurement strategy is either a geometric series (black), random sampling (red), or our AI-driven approach (blue). Shaded area above and below the plotted markers represents the standard deviation of the mean accuracy. Both axes use logarithmic scaling.}
\label{fig:sim_acc_evolution}
\end{figure}

      \textit{In silico} phase mapping experiments were seeded with a coarse grid of 49 points (Figure \ref{fig:AIdriven_example}A) and for each measurement strategy a total of 10 experiment campaigns were done for averaging. It was found that the accuracy of models predicted from AI-driven mapping (blue) surpassed that of both the geometric series (black) and random sampling (red) after only 1 additional measurement is done post seeding (Figure \ref{fig:sim_acc_evolution}). Generally, the difference in average model accuracy between the AI-driven and conventional mapping strategies increases as more measurements are done, such that after 5 \% of the total data is collected the accuracies are 98.9 \%, 95.9 \%, and 96.6 \% for the AI-driven, geometric series, and random sampling approaches, respectively. Additionally, there is a pronounced difference in relative change in accuracy after approximately 1\% of the total measurements possible are done as shown by the increasing slope for the AI-driven campaign accuracy as opposed to the decreasing slope for the geometric series accuracy. It is also notable that random sampling outperforms the geometric series after $\approx$2 \% of the data is collected, suggesting that homogeneous mapping approaches can benefit from being supplemented with random sampling for better exploration (\emph{e.g.}, when features are localized in regions that grids miss entirely). 

\begin{figure}
\centering
\includegraphics[width=\textwidth]{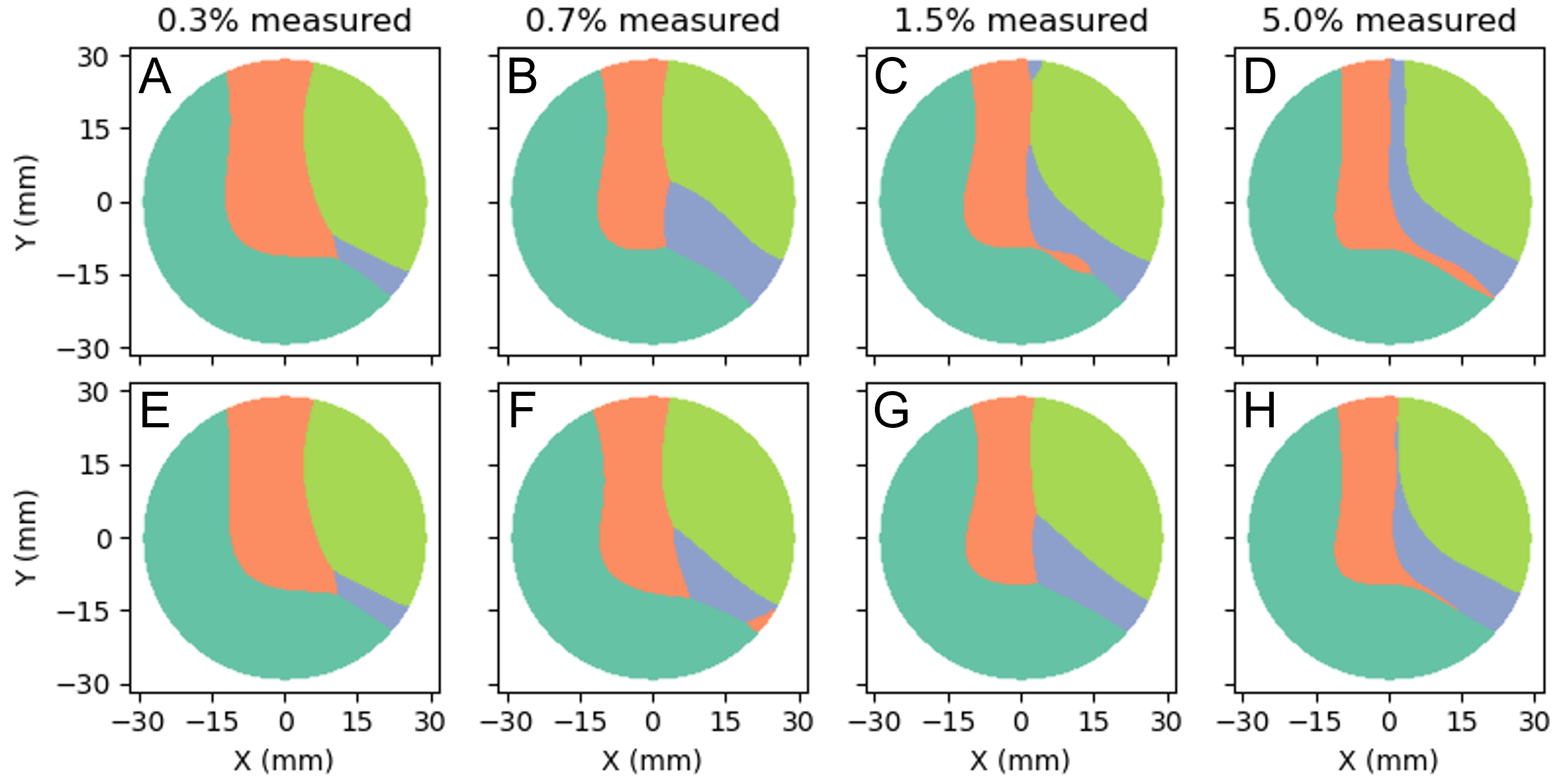}
\caption{Snapshots of simulated experiment campaign progress using (A-D) an AI-driven approach and (E-H) a conventional geometric series. Models of the simulated sample after collecting (A, E) 0.3 \%, (B, F) 0.7 \% (C, G) 1.5 \%, and (D, H) 5.0 \% of total possible measurements are predicted by a Gaussian process classifier trained after each measurement on measured coordinates and class labels.}
\label{fig:sim_camp_snapshots}
\end{figure}

      Though accuracy is an important metric for quantifying the performance of these mapping strategies, visualizing the predicted models for comparison to the ground truth informs what features are captured or missed. Snapshots of the models predicted after (0.3, 0.7, 1.5, and 5.0) \% of total possible measurements are done show that measurements from the AI-driven campaign (Figure \ref{fig:sim_camp_snapshots}A-D) enable significantly better predictions of both the size and shape of phase regions as well as their interfaces in comparison to those collected in a conventional grid (Figure \ref{fig:sim_camp_snapshots}E-H). Overlaying measurement locations on the model snapshots (Figure S16) reveals that the AI-driven mapping prioritizes measurements along phase boundaries and in regions with more rapid changes as a function of spatial distance, while limiting measurements in regions with smooth changes (\emph{e.g.}, within the teal and lime-green sections). The value of the measurements selected algorithmically is further highlighted in difference maps of these campaign snapshots (Figure S17) in which the incorrectly predicted coordinates are marked in black and show that phase boundaries are where the GP trained on measurements done in a grid most struggle to predict labels. While minor phase combinations (colored in pink and yellow in Figure \ref{fig:sim_ground_truth}) at the interface of major phase regions are not predicted by the GP from either set of measurements, these localized features are measured in the AI-driven campaign and missed in the conventional grid. This suggests that even when the modeling capabilities of a GP (defined by the kernel and hyperparameters) are not tuned to the features present, AI-driven mapping methods can still enable discovery beyond the capabilities of traditional approaches. Lastly, snapshots of the models predicted throughout a campaign, plus overlays of the measurements done (outlined pixels), acquisition function values (dark to light) and the next selected measurement (red Xs) are compiled into movies that allow the experiment campaign to be played back. Replaying these campaigns confirms the behavior observed in the snapshots (Figure \ref{fig:sim_camp_snapshots}, S16, S17) with the added benefit of visualizing the measurement order to establish that the AI-driven campaign first explores unknown regions, then prioritizes measurements along phase boundaries. In addition to developing the real-world framework for AI-driven, multi-modal studies, we have demonstrated through experiment campaigns conducted \textit{in silico} that AI-driven phase mapping approaches outperform both conventional grid strategies and random sampling.

\section{Discussion}

\subsection{Necessary infrastructure development.}
      The multi-modal, multi-beamline, and AI-driven experimental workflows discussed are enabled by several key technologies and institutional investments such as the building of a High Throughput Science Network\cite{Rakitin2022} and experiment orchestration using the Bluesky suite (Figure S18). Unifying themes include integrating industry-standard technologies to enhance the scale and reliability of the facility, and decentralized control of the beamlines to move beyond closed loop experimentation toward flexible and collaborative human-AI operating modes.\cite{Maffettone2023a,Konstantinova2022} Additionally, the principles of FAIR\cite{Wilkinson2016} (findability, accessibility, interoperability, and reusability) data that have been identified as vastly beneficial to both individual researchers and the broader scientific community, especially for curating labeled training data, are accounted for.
      
      One critical upgrade at NSLS-II was a network reprovisioning process that connected all beamlines as well as the accelerator to the rest of the BNL campus.\cite{Rakitin2022} This provides a centrally managed network for various instruments and devices on the experiment floor. It also allows the facility to connect to external resources for data storage and computing such as the Scientific Data and Computing Center at BNL that hosts JupyterHub\cite{Milligan2017,Rind2020}, a platform for authenticated, remote data access and analysis. Additionally, whether experiments are driven by AI agents or users, the firewalled communications between beamlines enabled by this upgrade are needed for safe and secure instrument controls.
      
      Another major infrastructure development necessary for this work is a modern, modular, and highly extensible suite of tools for experiment orchestration. The Bluesky suite, which was first developed at NSLS-II and is now being contributed to by the larger synchrotron science community (\emph{e.g.}, at Advanced Light Source, Canadian Light Source, Diamond Light Source, Australian Synchrotron), is an optimal solution. First, the Bluesky data model is flexible to handle the variety of data and metadata that are produced at a large-scale facility with diverse techniques and instrumentation. Second, Bluesky was developed in Python which makes it interoperable with a plethora of open-source libraries as well as with bespoke scientific tools. Third, it  has native data streaming capabilities that move beyond antiquated file storage systems that do not scale with modern data volumes and access routines while retaining compatibility with legacy systems. Bluesky is a collaborative project that is future looking, flexible, and supported by facility- and community-wide development, thus it avoids design decisions that could silo its usability to a single beamline or technique.
      
      Recently developed components of the Bluesky ecosystem that were particularly important for data streaming and asynchronous autonomous operating modes are Tiled and Bluesky Adaptive\cite{BrookhavenNationalLaboratory2025b}, respectively. Tiled provides secure search and access to data via a hypertext transfer protocol service (https) end point and a Python client library that integrates with data science libraries such as Xarray\cite{Hoyer2017} and pandas\cite{McKinney2010} for data slicing and sub selection. This service sits atop databases and file systems alike, and includes capabilities for querying (\emph{e.g.}, by beamline proposal number, datetime), as well as structured, chunk-wise access that is necessary for large data volumes. Bluesky adaptive provides an adaptive run engine for decentralized control of beamlines through a queue server that manages permissions and secure access by both users and AI agents. Virtualization of our workflow (Figure \ref{fig:multimodal_blockdiagram}) stands the separable components up as independent services, ensuring that an agent misbehaving can be restarted or reconfigured without bringing down the full workflow. Further, message bus services allow customization of which data streams are relevant to each agent and inform data streaming protocols through publish / subscribe methods. The synergy of these tools provides enhanced operational reliability, adaptive configuration controls, and asynchronous modalities for conducting experiments out of lockstep.
      
      Lastly, it is evident that simulated systems are needed for developing and testing both agents and workflows. The Bluesky suite enables this through automated metadata capture (\emph{e.g.}, motor positions, ring current, external device logging) that is essential for data-hungry ML/AI methods which greatly benefit from labeled data. Additionally, experiment replay capabilities are native in Bluesky due to the document model in which standard data structures are stored in memory rather than files. This facilitates replay because document consumers are indifferent to the source of the streamed data, whether it be from an acquisition process or data storage.

\subsection{Considerations of AI-driven methods.}
      The evolution of beamline operations from mostly manual to highly automated and now toward autonomous operating modes is analogous to the evolution of automobiles. For example, first-generation vehicles required manual steering and gear shifting, then later designs incorporated automatic shifting and power steering, and now assisted- and self-driving modes are enabled by Light Detection and Ranging (LiDAR) and computer-vision technologies.\cite{HighwaySafety} While the main utility of synchrotron beamlines and vehicles has remained mostly the same (materials characterization and personal travel, respectively), the user experience has been significantly enhanced by the incorporation of new technologies and quality-of-life improvements. The integration of AI with beamline operations is one of the current frontiers for synchrotron science.
      
      Access to a range of operating modes spanning collaborative human/AI control to fully autonomous (Table \ref{tab:operating_modes}) must be available because experimentation is constantly evolving and the appropriate level of automation (Table \ref{tab:automation_levels}) depends on the use-case and precedent for safe, reliable operations. Automation is defined as computer-driven control, while automated analysis is effectively a separate class of tools which in isolation do not pose a risk or added hazard to the experiment. If an automated analysis tool is used in an automated workflow, it must be considered in the risk assessment (\emph{e.g.}, is there validation of its output, can the automated analysis lead to an unsafe or unproductive use of beamtime). For example, in any operating mode in which AI agents are able to provide control instructions such as moving a motorized stage there must be engineering controls to prevent collisions whether through soft motor limits validated by a human or active visual monitoring systems. In addition to hazards, productive use of beamtime is imperative thus fully autonomous experiments are only appropriate when the systems are well understood. This includes the material's physicochemical properties (\emph{e.g.}, reactivity), sample environment (\emph{e.g.}, high temperature or pressure, gas flow), and precedent for automated analysis (\emph{e.g.}, successful feature identification and modeling) that can intelligently drive experiments. 

\begin{table}[hb]
  \centering
  \caption{Beamline operating modes that integrate AI and the roles of humans and AI.}
  \label{tab:operating_modes}
  \setlength{\arrayrulewidth}{0.4pt}
  \begin{tabularx}{\linewidth}{|C{0.18\linewidth}|C{0.18\linewidth}|C{0.18\linewidth}|Y|}
    \Xhline{1.2pt} 
    \textbf{Modality} & \textbf{Human Role} & \textbf{AI Role} & \textbf{Description} \\
    \hline
    Fully Autonomous & N/A & Full control & AI fully controls experiments based on real-time analysis; appropriate for well-understood systems. \\
    \hline
    Manual with AI assist & Full control & Advisory only & AI analyzes data and suggests next steps but does not control equipment. \\
    \hline
    Collaborative Control & Shared with AI & Shared with AI & Both AI and human experts provide instructions via a priority system. \\
    \Xhline{1.2pt} 
  \end{tabularx}
\end{table}

\begin{table}[t]
  \centering
  \caption{Levels of beamline automation and example tasks.}
  \label{tab:automation_levels}
  \setlength{\arrayrulewidth}{0.4pt}
  \begin{tabularx}{\linewidth}{|C{0.15\linewidth}|Y|C{0.42\linewidth}|}
    \Xhline{1.2pt}
    \textbf{Automation Level} & \textbf{Description} & \textbf{Example} \\
    \hline
    0 & Manual control only & Direct EPICS commands, hardware testing \\
    \hline
    1 & Task-specific scripts & Moving motors to set positions; simple repetitive tasks \\
    \hline
    2 & Scripts with feedback control & Automated beam alignment using signal optimization \\
    \hline
    3 & Goal-oriented automation with decision-making & Detect drift, take corrective actions, then resume measurements \\
    \Xhline{1.2pt}
  \end{tabularx}
\end{table}

      Levels of beamline automation are defined by the task or objective (Table \ref{tab:automation_levels}) from level 0 to 3. Fully manual control (level 0) is not frequently encountered by users, as this represents tasks like direct hardware control with EPICS commands which are abstracted away by Bluesky interface layers (Figure S18). Level 1 represents the conventional synchronous approach to experiment control in which task specific scripts (\emph{e.g.}, N motor movements followed by measurements in a loop) are manually configured and initialized, then executed at runtime. Improved experimental outcomes can be achieved by utilizing feedback within task-specific scripts in a pre-defined manner (level 2), such as maximizing a signal by tuning motor positions within limits. The highest level of automation (level 3) includes goal-specific operations that require a degree of automated decision making to go beyond pre-defined and potentially ill-conditioned bounds inherent to level 2.
      
      Understanding which level of automation is appropriate for the task at hand is necessary for the successful use of AI in facility operations. For example, in the case of a MX beamline that requires precise sample handling for routine data collection to meet a demanding sample queue (ideally measuring tens to hundreds of samples / day), a highly automated workflow (level 3) can improve reliability and reproducibility through standardization and anomaly detection. In contrast, studying a temperature-dependent gas-flow synthesis reaction often requires a higher degree of user intervention due to the engineering complexity and risk factors introduced by the electrical and chemical hazards (level 1 - 2). The need for varying degrees of autonomy applies beyond measurements at light and neutron sources, as an extensible multi-modal experiment orchestration platform must be compatible with instruments at other facilities such as nanoscale research centers, and data from simulations (\emph{e.g.}, DFT) or historical sources (\emph{e.g.}, previous measurements, literature-mined data). 

\subsection{Data abstraction in autonomous experimentation - raw data \emph{vs.} scientific products.}
      Thus far, we have discussed AI-driven approaches that use measured data such as 1D XRD patterns and XAFS spectra as input. While it is evident that these algorithms can recognize patterns in the data that humans are unable to discern by simply visualizing the data, physics-based scientific analyses such as Rietveld refinement\cite{Rietveld1969} and XAFS modeling\cite{Ravel2016} can be used to extract detailed structural and chemical insights (\emph{e.g.}, unit cell parameters, phase fractions, coordination environment) that are not highlighted in the raw data. For example, a clustering algorithm like K-means operates on hundreds or thousands of data points per measurement and may not be sensitive to subtle yet systematic changes in diffraction peak intensities from chemical substitution. However, modeling the source of these seemingly inconsequential changes (\emph{e.g.}, by refining atomic site occupancies) can reveal important structural differences and provide a more informative descriptor for resolving heterogeneities within a dataset. 
      
\begin{figure}
\centering
\includegraphics[width=\textwidth]{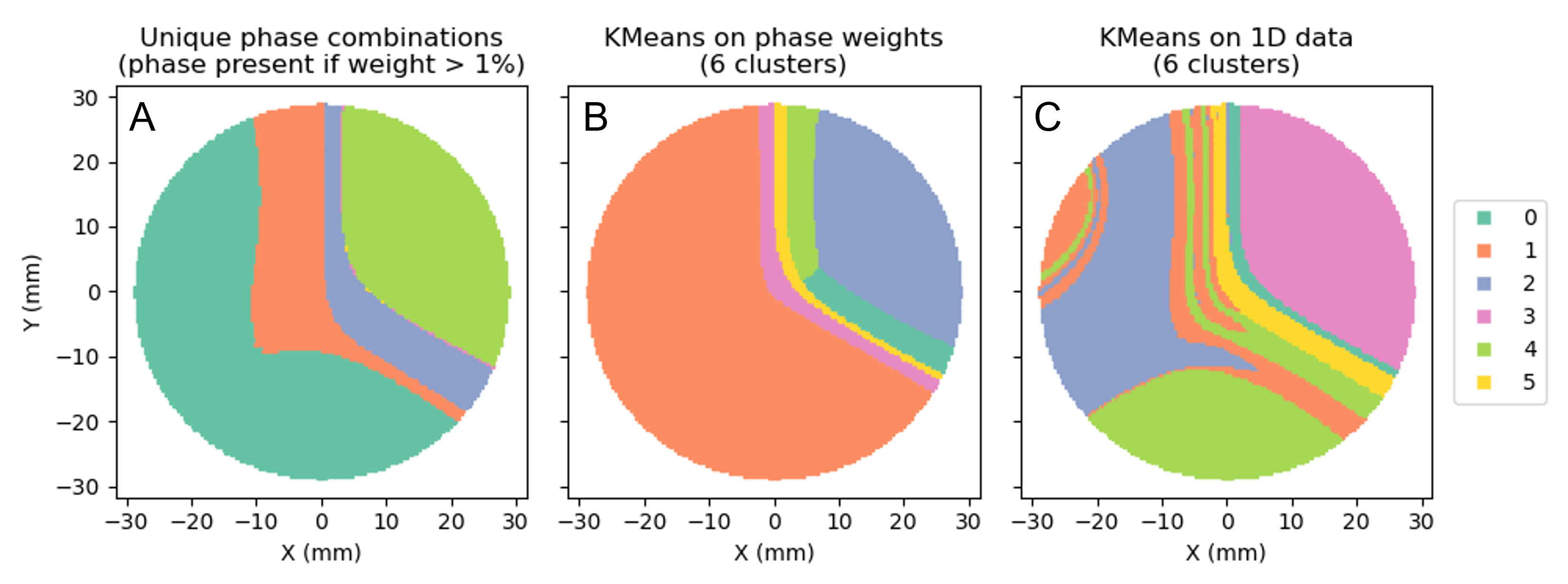}
\caption{Comparison of labeling approaches for the simulated Al-Li-Fe oxide sample, where the marker colors teal, orange, purple, pink, lime green, and yellow each represent a unique label. A) Labels assigned based on unique combination of phases present. B) Labels from K-means clustering with 6 clusters when the input data are phase weights. C) Labels from K-means clustering with 6 clusters when the input data are 1D XRD patterns.}
\label{fig:data_abstraction_comp}
\end{figure}

      The optimal level of data reduction for input to AI-driven approaches will depend on the scientific questions that researchers are trying to answer and the nature of the experiment and measurement. As an example, maps of the simulated Al-Li-Fe oxide sample (Figure \ref{fig:data_abstraction_comp}) differ drastically when labels for each coordinate are A) assigned according to the unique combination of phases present, B) derived from K-means clustering on phase weights, and C) derived from K-means clustering on 1D diffraction data. The measurements prioritized by AI agents using these labels to steer an experiment would clearly be unique from one another because the size, shape, and interfaces of these clusters differ greatly. While identifying the optimal data type for input to AI agents targeting a given objective is beyond the scope of the present work, it is evident that scientific products extracted by traditional physical modeling approaches provide sensitivity to discrete features that would require careful feature engineering in ML and AI-based approaches. Further, the features that provide the most valuable insight are often not known in advance, especially for novel materials. This highlights the advantage of our plug-and-play capability for AI agents that allows various analyses to be done in tandem and for the input to decision-making agents to be modified on-the-fly through accessible configuration controls.

\section{Conclusions}
      The modular framework presented here establishes a practical path toward autonomous, yet still collaborative, synchrotron experiments across multiple instruments. We report a world's first achievement enabled by the Bluesky suite and data infrastructure at NSLS-II: real-time, synchronous, dynamic orchestration of complementary XRD and XAFS measurements on multiple beamlines by an ensemble of AI agents with human-in-the-loop. By integrating AI agents for data reduction, analysis, and Bayesian optimization-based decision-making through Bluesky adaptive, the workflow supports both synchronous and asynchronous multi-modal studies while preserving facility safety considerations through the queue server. Researchers will be able to select operating modes ranging from manual to autonomous, ensuring that human expertise remains central when needed.
      
      Deployment of this workflow for multi-modal mapping of Al-Ni-Pt combinatorial libraries demonstrated successful inter-beamline communication, on-the-fly data processing, and adaptive switching between autonomous and grid-based acquisition strategies. Simulated campaigns on a digital twin confirmed that AI-driven mapping approaches surpass conventional geometric grids in terms of resolving phase boundaries and measuring highly localized minor components. These results suggest that facility throughput can be increased substantially for mappable samples where the scientific information is spatially heterogeneous.
      
      The framework is interoperable with open-source tools in the Python ecosystem, permits plug-and-play extension with traditional physics-based or data-driven agents, and enables data practices in line with FAIR data principles. Isolated deployment of agents as services, combined with offline testing in the digital twin, lowers the barrier for community contributions while minimizing risk to beamline operations. Benchmarking of agent performance, expansion to additional modalities, and development of governance models for autonomous control will be essential next steps. The infrastructure reported here provides a foundation on which multi-modal, multi-beamline, AI-assisted experiments can become routine practice across user facilities. Looking forward, we envisage a collaborative human-AI scientific ecosystem in which heterogeneous data streams from scattering, spectroscopy, imaging, and simulation converge with materials acceleration platforms such as inline synthesis to accelerate materials discovery and scientific advancements as a whole. 

\section{Experimental}

\subsection{Thin film preparation.}
      Two nominally identical films with a composition gradient were deposited using combinatorial co-sputtering\cite{Joress2022} using (3) 2 in.\@ (50 mm) elemental targets: Al, Ni, and Pt ($\geq$ 99 \% purity). 
      Films were sputtered onto 75 mm fused silica wafers, a small portion of which was masked to allow for a background diffraction pattern to be measured. The magnetron sputter guns were spaced $120^\circ$ from each other and aimed away from the center of the wafer in order to create a lateral variation in thin film composition. The base pressure of the chamber was 10 $\mu$Pa and sputtering was done under an Ar atmosphere of 930 mPa (7 mTorr). During deposition, the substrates were heated to $100^\circ$C. For sputtering, the Al, Ni, and Pt guns were run at 95 W, 40 W, and 30 W, respectively. The Al gun was run for 30 s prior to the start of co-deposition, with the substrate rotating, to create a bond layer to help with film adhesion. Each film was sputtered for 90 min.

\subsection{Synchrotron measurements.}
      Powder diffraction measurements were performed on the Pair-Distribution Function (PDF) beamline 28-ID-1 at NSLS-II using a Dectris Pilatus31 2M-L CdTe detector at 74.46 keV ($\lambda = 0.1665$\,\AA). Each measurement was collected for 30 seconds, then integrated and background subtracted using the standard beamline data processing software. Both the raw detector images and reduced diffraction patterns were stored in the Databroker framework, and immediately accessible using Tiled, with metadata including measurement time, wafer position (absolute and relative), beam current, detector positions, and calibration information. The PDF beamline was controlled using the Bluesky queue server infrastructure, with users and AI agents (when enabled) able to add plans to the queue during the measurement. X-ray absorption spectroscopy measurements were performed on the Beamline for Materials Measurement (BMM) beamline 6-BM at NSLS-II in fluorescence mode using a 4-channel Hitachi Vortex\textregistered-ME4 silicon drift detector with Xpress3 for detector readout and deadtime correction. 

\subsection{Diffraction analysis.}
      Powder diffraction patterns were analyzed on-the-fly at the beamline using GSAS-II\cite{Toby2013} for Rietveld refinements. Post-experiment TOPAS\cite{Coelho2018} (Bruker AXS, version 7) was used for Rietveld and Pawley fitting. In both software packages the instrument profile function was analytically determined by fitting a pseudo-Voigt function to the peak profiles in data collected on a NIST Si standard (SRM 640f, \emph{a}=5.43114 \AA) in the same instrument configuration. A 3-term polynomial was fit to the standard data to correct for minor peak position offsets due to parallax effects.

\subsection{Computational framework and agent design.}
      The computational framework and AI agents were developed entirely in the python ecosystem, ensuring ease of use and extensibility. Agents inherit basic methods from a generic parent agent class for retrieving and storing data, such that agent configuration, state, and outputs can be queried in addition to the measured data. 

\subsection{Digital twin for diffraction measurements.}

      A 2D wafer representative of a combinatorial library was simulated from a calculated diffraction dataset for the Al-Li-Fe oxide system that was designed for benchmarking materials discovery and autonomous phase mapping.\cite{Chen2021,Stanev2018,LeBras2011,LeBras2014}
      To accomplish this, we designed a flexible simulator for ``mappable'' systems. First, a discretized, mesh grid of 2D coordinates is created. Then, an arbitrary number of points are selected to represent the nozzle streams in the deposition process, and their positions can either be manually entered or calculated using a distance parameter (0 to 100 where 0 = origin, 100 = edge) that distributes the points equidistant from each other along the edge of the shape - in this case we are simulating a circular wafer so the points are distributed as a function of the circumference of a circle with a radius equal to the distance parameter. The elemental composition at these fixed points is inputted (\emph{e.g.}, [0.6, 0.2, 0.2]), then the composition at all other coordinates is calculated using the Euclidean distance between a given point and the fixed points with a Gaussian smoothing function. Phase weights and a diffraction pattern corresponding to the elemental composition at each position are interpolated from the known relationships between elements, phases, and diffraction patterns in the calculated phase diagram, including relative changes in diffraction peak positions and intensities from chemical substitution. If the wafer shape is set to be a circle (as was done in this case), then points outside the radius of the circle are masked such that the elemental weights, phase weights, and diffraction patterns are NaNs. An Xarray\cite{Hoyer2017} dataset is assembled from the various multi-dimensional arrays, providing a compact data structure that captures indexed coordinates, phase weights, and diffraction patterns along with metadata including the fixed composition values and coordinates, and spatial resolution.
      
      Python code was developed to enable experiments \textit{in silico} on both simulated and previously measured samples. For the measurement campaigns discussed (\emph{i.e.}, geometric series, random sampling, AI-driven) a Gaussian process classifier\cite{Rasmussen2005} (GPC, scikit-learn\cite{FabianPedregosa2011} v1.6.0) was used for model generation. A GPC was trained on coordinates where measurements were done and the corresponding labels representing the unique combinations of phases present, then the labels at all unmeasured coordinates were predicted. The model of the sample was constructed from the labels at all points, and the class probabilities of the GPC prediction were used to calculate the Shannon entropy which we refer to as uncertainty. An edge detection algorithm then compares the label at a given position to the labels of neighboring coordinates and assigns an edge value, such that coordinates with neighbors sharing the same label are given an edge value of 0 and those that have a neighbor with a different label are given an edge value of 1. This edge characteristic is then combined with the GPC uncertainty, and a cumulative distribution function is constructed from the edge-weighted uncertainty. In the case of AI-driven campaigns measurement suggestions are sampled from this function. 

\section{Author Contributions}
PMM, BR, TC, SC, SW, and DO conceptualized the project. HJ prepared the Al-Ni-Pt thin films. AAC, PMM, BR, TC, and DO conducted synchrotron XRD and XAFS measurements. AAC analyzed the Al-Ni-Pt diffraction data and BR analyzed the XAFS data. PMM implemented Bluesky Adaptive workflows that enable synchronous measurements. PMM developed the AI agent architecture and agents for Bayesian optimization driven phase mapping. AAC developed agents for peak fitting and Rietveld refinements. AAC simulated the Al-Li-Fe sample and measurement campaigns. AAC drafted the manuscript with input from all authors. AAC, DO, and BR edited the manuscript with input from all authors.

\section{Abbreviations}
XRD, X-ray Diffraction; PDF, Pair-Distribution Function; XAFS, X-ray-Absorption Fine-Structure; XANES, X-ray Absorption Near-Edge Structure; EXAFS, Extended X-ray Absorption Fine Structure; ML, Machine Learning; AI, Artificial Intelligence; GP, Gaussian Process; BMM, Beamline for Materials Measurements; NSLS-II, National Synchrotron Light Source II.

\section{Acknowledgment}
This research used resources of the National Synchrotron Light Source II, a U.S. Department of Energy (DOE) Office of Science User Facility operated for the DOE Office of Science by Brookhaven National Laboratory under Contract No. DE-SC0012704.  

\section{Declaration of Interests}
The authors declare no competing interests, financial or otherwise.

\section{Data Statement}
The data that support the findings of this study are available from the corresponding author upon reasonable request. Example code and datasets are open-source under the BSD 3-clause license and are available at \url{https://github.com/AdamCorrao/pub-Corrao_2025_09}\cite{BrookhavenNationalLaboratory2025c}. 

\printbibliography

\end{document}